\documentclass[useAMS, onecolumn]{mn2e}

\pdfoutput=1
\usepackage{epsfig}
\usepackage{amsmath}
\usepackage{amssymb}
\usepackage{appendix}
\def\be{\begin{equation}}
\def\ee{\end{equation}}
\def\ba{\begin{eqnarray}}
\def\ea{\end{eqnarray}}
\def\go{\mathrel{\raise.3ex\hbox{$>$}\mkern-14mu
             \lower0.6ex\hbox{$\sim$}}}
\def\lo{\mathrel{\raise.3ex\hbox{$<$}\mkern-14mu
             \lower0.6ex\hbox{$\sim$}}}
\def\bxi{{\mbox{\boldmath $\xi$}}}

\def\bnab{{\mbox{\boldmath $\nabla$}}}

\def\bomega{{\bar\omega_\alpha}}

\begin{document}

\title[Tides in Rotating White Dwarfs]
{Dynamical Tides in Compact White Dwarf Binaries: Influence of Rotation}
\author[J. Fuller and D. Lai]
{Jim Fuller$^{1,2}$\thanks{Email:
jfuller@caltech.edu; dong@astro.cornell.edu}
and Dong Lai$^3$\\
\\$^1$TAPIR, Mailcode 350-17, California Institute of Technology, Pasadena, CA 91125, USA
\\$^2$Kavli Institute for Theoretical Physics, Kohn Hall, University of California, Santa Barbara, CA 93106, USA
\\$^3$Center for Space Research, Department of Astronomy, Cornell University, Ithaca, NY 14853, USA}

\label{firstpage}
\maketitle

\begin{abstract}

Tidal interactions play an important role in the evolution and ultimate fate of compact white dwarf (WD) binaries. Not only do tides affect the pre-merger state (such as temperature and rotation rate) of the WDs, but they may also determine which systems merge and which undergo stable mass transfer. In this paper, we attempt to quantify the effects of rotation on tidal angular momentum transport in binary stars, with specific calculations applied to WD stellar models. We incorporate the effect of rotation using the traditional approximation, in which the dynamically excited gravity waves within the WDs are transformed into gravito-inertial Hough waves. The Coriolis force has only a minor effect on prograde gravity waves, and previous results predicting the tidal spin-up and heating of inspiraling WDs are not significantly modified. However, rotation strongly alters retrograde gravity waves and inertial waves,
with important consequences for the tidal spin-down of accreting WDs. We  identify new dynamical tidal forcing terms that arise from a proper separation of the equilibrium and dynamical tide components; these new forcing terms are very important for systems near synchronous rotation. Additionally, we discuss the impact of Stokes drift currents on the wave angular momentum flux. Finally, we speculate on how tidal interactions will affect super-synchronously rotating WDs in accreting systems.

\end{abstract}


\section{Introduction}

Merging and mass-transferring white dwarfs (WDs) are responsible for
the creation of a variety of exotic astrophysical systems and
events. Among these 
are isolated sdB/sdO stars (likely created
by the merger of two He WDs; Saio \& Jeffery 2000; Han et al. 2002;
Heber 2009), R CrB stars (likely created by the merger of a CO and He
WD; Webbink 1984; Iben et al. 1996; Jeffery et al. 2011; Clayton 2012), 
AM CVn binaries (possibly created by stable mass transfer from a He WD
onto a CO WD; Warner 1995; Tutukov \& Yungelson 1996; Han \& Webbink
1999; Nelemans et al. 2001a), and possibly 
high-mass neutron stars and magnetars (created by the
accretion-induced collapse of merging CO WDs or accreting O-Ne-Mg WDs;
Saio \& Nomoto 1985; Nomoto \& Kondo 1991; Saio \& Nomoto 2004).
More importantly, merging or accreting
CO WDs may be the progenitors of type Ia supernovae (Iben \& Tutukov
1984; Webbink 1984, Di Stefano 2010; Gilfanov \& Bogdan 2010; Maoz et
al. 2010; Li et al. 2011; Bloom et al. 2012; Gonzalez Hernandez et
al. 2012; Schaefer \& Pagnotta 2012), although the pre-supernova
conditions of the WD are still a subject of 
intense investigation.

Compact WD binaries (with periods in the range of minutes to hours)
are also being discovered at an accelerating pace (e.g. Mullally et
al. 2009; Kulkarni \& van Kerkwijk 2010; Steinfadt et al. 2010; Brown
et al. 2012; Kilic et al. 2012; Kilic et al. 2013, see Marsh 2011 for
a review). These systems will begin 
mass transfer 
within a Hubble time, although it is not always clear which observed
WD systems will result in which phenomena. Recent simulations (e.g.,
Segretain et al.~1997; Yoon et al.~2007; Loren-Aguilar et al.~2009;
van Kerkwijk et al.~2010; Dan et al.~2012; Raskin et al.~2012, Dan et
al. 2013) have shed new light on the outcome of WD mergers, however,
the initial conditions of these simulations are predicated upon the
tidal effects that precede (and may ultimately determine the stability
of) the mass transfer process. It is thus crucial to understand how
tides operate up to the moment of Roche-lobe overflow.

It is also important to quantify the rate of tidal dissipation once
mass transfer begins. Previous works (Nelemans et al. 2001b, Marsh et
al. 2004) have shown that the strength of tidal torques may be
important in determining the stability of mass transfer. Low mass
ratio systems with $M_2/M_1 \lesssim 1/4$ are expected to have stable
mass transfer, while high mass ratio systems with $M_2/M_1 \gtrsim
2/3$ are expected to be unstable and merge. The fates of intermediate
mass ratio systems with $M_2/M_1 \sim 1/2$ (such as the recently
discovered 12.75 minute binary, Brown et al. 2011) are less
certain. The stability of mass transfer in these systems may depend on
the rate at which tidal torques can transfer rotational angular
momentum from the accretor back to the orbit.

This paper is the fifth in a series (see Fuller \& Lai
2011,2012a,2012b,2013, hereafter Papers I-IV) dedicated to the physics
of tidal interactions and their observational consequences in compact
WD binaries. In these previous works, we examined the tidal
excitation, propagation, and dissipation of gravity waves (g-waves)
that are restored by buoyancy forces within CO and He WDs. The waves
dissipate energy and angular momentum, leading to the rotational
synchronization and tidal heating of WDs in compact systems. We found
that gravity waves are excited at compositional gradients within the
WDs and propagate toward the surface where they become highly
non-linear, causing them to break and locally deposit their angular
momentum. This results in strong tidal heating in the outer layers of
the WD, and causes the WDs to be spun up from the outside in.

Tides in white dwarfs have also been examined in several previous
works using parameterized tidal dissipation laws (Iben et al. 1998,
Piro 2011, Dall'Osso \& Rossi 2013) and in calculations of the tidal
excitation of oscillation modes (Willems et al. 2010, Valsecchi et
al. 2012). Encouragingly, the detailed study of WD tides by Burkart et
al. 2013 utilizes techniques comparable to 
Papers I and II, and reaches generally similar conclusions.

However, these works (and all previous works that we know of) have
ignored the influence of rotation on the wave dynamics. While this
approximation works for slowly spinning WDs at the beginning of the
tidal spin up process, it breaks down for WDs nearing synchronous
rotation in which the tidal forcing frequency becomes comparable to
the spin frequency. 
The approximation is even worse 
for WDs being spun up to super-synchronous rotation via mass transfer.

In this study, we attempt to quantify the effect of the Coriolis force
on the efficiency of tidal angular momentum transport. To do this, we
employ the so-called \textquotedblleft traditional" approximation
frequently used to study the influence of rotation on gravity waves
(see e.g., Chapman \& Lindzen 1970, Bildsten et al. 1996, Lee \& Saio 1997, Townsend 2003). The Coriolis force
modifies the angular dependence of the gravity waves, confining them
to the equatorial region of the star. This affects their overlap with
the (predominantly quadrupolar) tidal potential that excites
them. Moreover, the Coriolis force modifies the radial wavelength of
gravity waves, which in turn affects the amplitude to which they are
excited. Finally, the Coriolis force allows for the existence of
inertial and Rossby waves (r-waves), which are especially important in
super-synchronously rotating WDs in which tidal dissipation occurs
through the excitation of retrograde waves (i.e., the wave pattern
propagates opposite to the direction of the spin in a frame
co-rotating with the WD).

This paper is organized as follows. In Section \ref{traditional}, we
review the influence of rotation under the traditional approximation,
and we examine its effect on tidally excited gravito-inertial
waves. In Section \ref{tidal}, we investigate the 
effects of rotation on the excitation of the dynamical tide and
discuss how the dynamical tide transports energy angular momentum. In
Section \ref{torque}, we present our numerical results of tidal
dissipation rates in rotating WDs and analyze its effect on the
orbital and spin evolution of compact WD binaries. Finally, in Section
\ref{conc}, we discuss the implications of our results, and how they
may affect WDs in various types of compact binary systems.

\section{Influence of Rotation in the Traditional Approximation}
\label{traditional}

The influence rotation on stellar oscillations has been examined in
many previous works, and we choose to follow along the lines of
Bildsten et al. 1996 and Lee \& Saio 1997 (although note the sign
difference in their definition of $\omega$ and $m$). Here we review
the basic physics in order to understand how rotation will affect
tidally excited waves in WDs. The perturbed linearized momentum
equation in the rotating frame of a spherical star may be written as
\be
\label{momeq}
-\rho \omega^2 \bxi = - \bnab \delta P - \rho \bnab U - g \delta \rho {\hat {\bf r}} + 2 i \rho \omega {\bf \Omega}_s \times \bxi.
\ee
Here, $\bxi$ is the Lagrangian displacement vector, $\delta P$ is the
Eulerian pressure perturbation, ${\bf \Omega}_s$ is the star's spin
vector, $U$ is the tidal potential of the companion star, and the
other quantities have their usual meaning. We have assumed a time
dependence $\bxi \propto e^{-i\omega t}$, and we have adopted the
Cowling approximation (i.e., we have ignored the perturbation to the star's
gravitational potential, which is a good approximation for the gravity
waves of interest). Additionally, we have ignored the centrifugal
force and stellar oblateness, which is a good approximation as long as
the stellar spin is much smaller than break-up, i.e., when $\Omega_s
\ll \Omega_{\rm dyn} = \sqrt{GM/R^3}$. Finally, we assume that the
star rotates rigidly along the $z$-axis such that ${\bf
  \Omega}_s=\Omega_s {\hat {\bf z}}$, with the spin vector aligned
with the orbital angular momentum.

For adiabatic oscillations, the perturbed momentum equations become
\be
\label{xir1}
-\rho \omega^2 \xi_r = -\frac{\partial}{\partial r} \Big( \delta P + \rho U \Big) + U \frac{\partial \rho}{\partial r} - \frac{g}{c_s^2} \delta P - \rho N^2 \xi_r - 2i\Omega_s \omega \rho \xi_\phi \sin\theta,
\ee
\be
\label{xit1}
-\rho \omega^2 \xi_\theta = -\frac{1}{r} \frac{\partial}{\partial \theta} \Big( \delta P + \rho U \Big) - 2i\Omega_s \omega \rho \xi_\phi \cos\theta,
\ee
\be
\label{xip1}
-\rho \omega^2 \xi_\phi = -\frac{1}{r \sin \theta} \frac{\partial}{\partial \phi} \Big( \delta P + \rho U \Big) + 2i\Omega_s \omega \rho \big(\xi_\theta \cos \theta + \xi_r \sin \theta \big).
\ee
Here we have used the adiabatic relation 
\be
\label{drho}
\delta \rho = \frac{1}{c_s^2} \delta P + \frac{\rho N^2}{g} \xi_r,
\ee
where $c_s$ is the sound speed and $N$ is the Brunt-Vaisala frequency. Additionally, the continuity equation implies
\be
\label{cont}
\delta \rho + \frac{1}{r^2} \frac{\partial}{\partial r} \big( \rho r^2 \xi_r \big) + \rho \bnab_\perp \cdot \bxi_\perp = 0,
\ee
where the $\perp$ subscript denotes the horizontal part of each vector. 

In the traditional approximation, the last term in equations
\ref{xir1} and \ref{xip1} are ignored. The traditional approximation
works well for gravity waves, which in the non-rotating WKB limit (and
with $l$ and $m$ of order unity) have $\xi_\theta \sim \xi_\phi \sim
(N/\omega) \xi_r$. Then, in regions in the star where $N \gg \omega$,
the last terms in equations \ref{xir1} and \ref{xip1} become
negligible and may be discarded. Equation \ref{xir1} then separates
from equations \ref{xit1} and \ref{xip1}, and the angular dependence
of the waves separates from the radial dependence.

The traditional approximation will be valid for the dynamical tide of
a rotating WD, which is comprised mainly of rotationally modified
gravito-inertial waves in the outer layers of the star where $N \gg
\omega$ (see Paper II). However, the traditional approximation does
not apply to the equilibrium tide, which is the hydrostatic response
of the star to the tidal potential, and which is dominated by
fundamental modes. To circumvent this problem, we decompose the
response of the star into equilibrium and dynamical components:
\be
\label{xidyn}
\bxi = \bxi^{\rm eq} + \bxi^{\rm dyn}.
\ee
We solve for the equilibrium tide $\bxi^{\rm eq}$ and subtract it from
equations \ref{xir1}-\ref{cont}, so that we can solve for $\bxi^{\rm
  dyn}$ using the traditional approximation. We detail this procedure
in Appendix \ref{eqtide}.

\subsection{Hough Functions}
\label{Hough}

The angular dependence of $\bxi^{\rm dyn}$ is found by solving Laplace's tidal equation:
\be
\label{lap}
\mathcal{L}(H_{k}) = - \lambda_k H_{k},
\ee
where $\lambda_k$ is an angular eigenvalue, $H_{k}$ is its associated
eigenfunction (the Hough function), and the operator $\mathcal{L}$ is
\be
\label{lop}
\mathcal{L} = \frac{\partial}{\partial \mu} \bigg( \frac{1-\mu^2}{1-q^2\mu^2} \frac{\partial}{\partial \mu} \bigg) - \frac{m^2}{(1-\mu^2)(1-q^2\mu^2)} + \frac{qm(1+q^2\mu^2)}{(1-q^2\mu^2)^2}.
\ee
Here, $\mu=\cos\theta$, and $m$ is the azimuthal number of the wave
such that $\bxi \propto e^{i m \phi}$. In this convention, prograde
waves (in the rotating frame) have $m>0$ and retrograde waves have
$m<0$. The parameter $q$ that determines the behavior of the Hough
functions is defined as
\be
\label{q}
q = \frac{2 \Omega_s}{\omega}.
\ee
Rotation becomes important for $q \gtrsim 1$, and it is easily
verified that the solutions to equation \ref{lap} converge to the
associated Legendre polynomials as $q \rightarrow 0$.

As discussed in Lee \& Saio (1997), the angular dependence of the
perturbed variables in the traditional approximation is
\be
\label{xirang}
\xi_r(r,\theta,\phi,t) = \xi_r(r) H_{k}(\mu) e^{i m \phi - i \omega t},
\ee
whereas 
\be
\label{xitang}
\xi_\theta(r,\theta,\phi,t) = \xi_\perp(r) \frac{1}{(1-\mu^2 q^2)\sqrt{1-\mu^2}} \bigg[-m q \mu - (1-\mu^2) \frac{\partial}{\partial \mu} \bigg] H_{k}(\mu) e^{i m \phi - i \omega t},
\ee
and
\be
\label{xipang}
\xi_\phi(r,\theta,\phi,t) = \xi_\perp(r) \frac{i}{(1-\mu^2 q^2)\sqrt{1-\mu^2}} \bigg[m  + q \mu (1-\mu^2) \frac{\partial}{\partial \mu} \bigg] H_{k}(\mu) e^{i m \phi - i \omega t}.
\ee
Here, $\xi_\perp(r) \equiv \delta P(r)/(\rho r \omega^2)$. We have adopted the indexing notation of Lee \& Saio (1997) to
identify the Hough functions. The $k$ subscript specifies the branch
on which the Hough function lies, and (for $k \geq 0$) indicates the
number of nodes in the angular eigenfunction in the non-rotating ($q
\rightarrow 0$) limit. Thus, the $k=0$ modes correspond to $l=|m|$
modes, $k=1$ corresponds to $l=|m|+1$, etc.

For $q<1$, equation \ref{lap} is a Sturm-Liouville problem, and 
each solution can be traced back to an associated Legendre
polynomial at $q=0$. For $q>1$, equation \ref{lap} is no longer a
Sturm-Liouville problem, and additional solutions exist. These
solutions correspond to r-modes and inertial modes, which only exist
in the range $\omega < 2 \Omega_s$, i.e., $q>1$. Lee \& Saio (1997)
label these solutions with negative values of $k$, and we adopt the same
convention here. However, we note that the $k=-2$ r-mode has no nodes
in its angular eigenfunction, similar to the $k=0$ mode. For this
reason, it will be the most important retrograde mode in tidal
processes for $q\gg1$.

We solve equation \ref{lap} for $m=\pm2$ in the range $10^{-2} < q <
10^2$ using relaxation techniques. This method avoids many of the
problems associated with singularities discussed by Bildsten et
al.~(1996) and Lee \& Saio (1997). We normalize the Hough functions in
the same manner as spherical harmonics:
\be
\label{norm}
\int\! dA\, |H_{k}(\mu)|^2 = 1,
\ee
with the integral extending over the surface of a sphere,
$\int\!dA =2\pi\int_{-1}^1 d\mu$.  We limit our analysis to $m=\pm 2$
waves because they couple most strongly to the tidal potential,
which is dominated by $l=|m|=2$ components.

\begin{figure*}
\begin{centering}
\includegraphics[scale=.4]{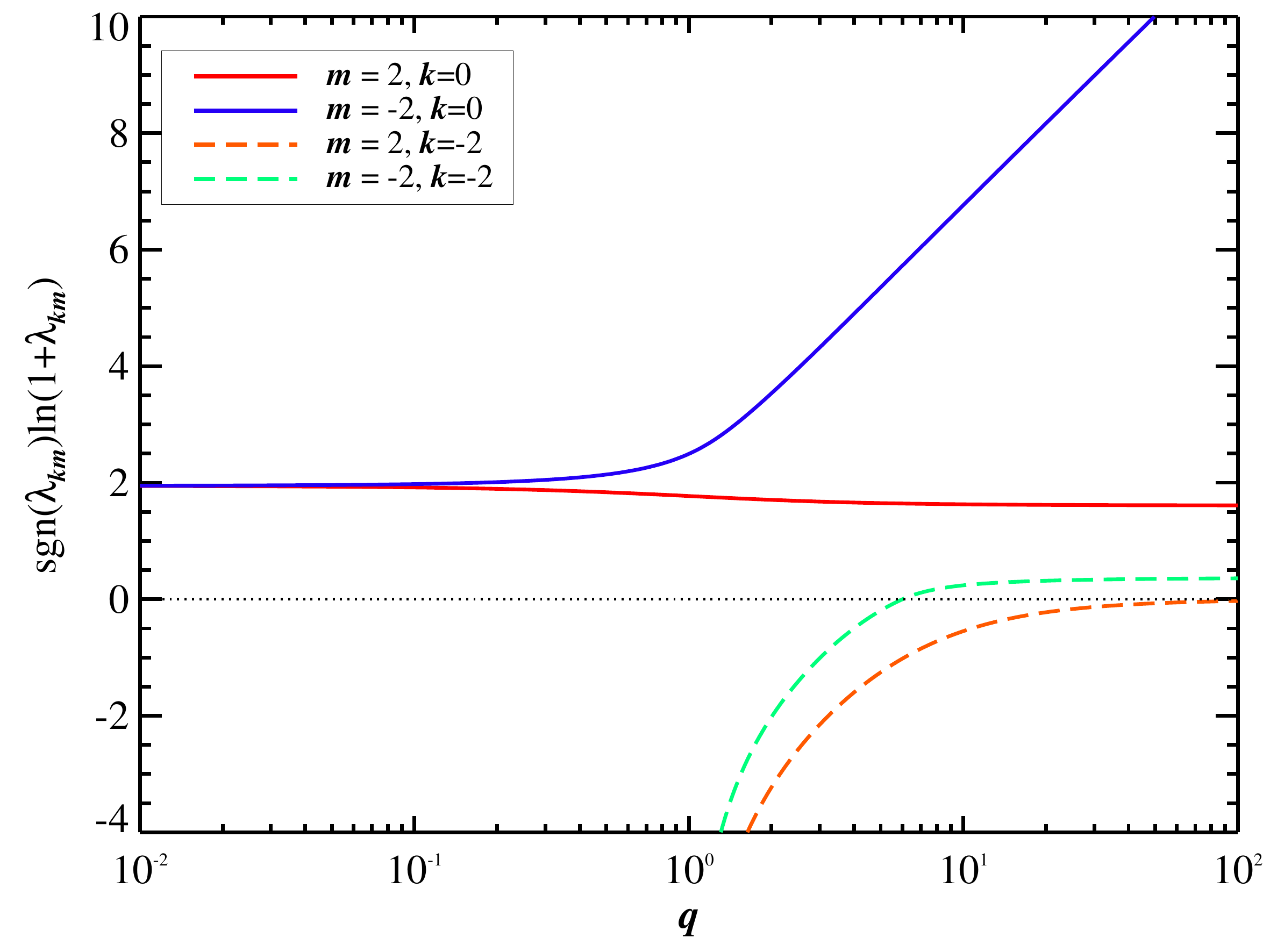}
\caption{\label{wdrotqlambda} Eigenvalue $\lambda_{k}$ of equation \ref{lap} as a function of the spin parameter $q=2 \Omega_s/\omega$, for a few different branches of solutions relevant to tidal dynamics. For comparison, we construct this plot in a similar fashion to Figure 1 of Lee \& Saio 1997. The horizontal dashed line denotes $\lambda_k = 0$, below which the traditional approximation breaks down in the bulk of the radiative interior of a WD. }
\end{centering}
\end{figure*}

Figure \ref{wdrotqlambda} shows a plot of the eigenvalues
$\lambda_{k}$ for the $k=0$ and $k=-2$ branches.\footnote{Odd values
  of $k$ are anti-symmetric about the equator and cannot be excited by
  a circular, aligned orbit. The $k>0$ branches couple most strongly
  with the $l>2$ components of the tidal potential, which are small and
  can be ignored.} For $q \ll 1$, the $k=0$ solutions converge to
$\lambda_k = |m|(|m|+1)$, as expected. For $q \gg 1$, the $m=2$, $k=0$
solution asymptotes to $\lambda_k = m^2$ while the $m=-2$ mode
increases proportional to $q^2$ as discussed in Bildsten et
al. 1996. As $q$ increases, the eigenfunctions $H_{k}$ become
localized near the equator, especially for the $m=-2$ solution.

The $k=-2$ solutions emerge for $q>1$. Their eigenvalues $\lambda_{k}$
asymptote to negative infinity at $q=1$ and become positive at $q =
(|m| + |k|)(|m|+|k+1|)/|m| = 6$ for $m=-2$, $k=-2$. The value of
$\lambda_{k}$ remains below zero at all values of $q$ for $m=2$, $k<0$ modes. Physically this is significant because waves in the WKB limit
follow the dispersion relation
\be
\label{gdisp}
k_r^2 = \frac{(L_{k}^2 - \omega^2)(N^2-\omega^2)}{\omega^2 c_s^2},
\ee
where $k_r$ is the radial wavenumber and 
\be
\label{lamb}
L_{k}^2 = \frac{\lambda_{k} c_s^2}{r^2}
\ee
is the Lamb frequency squared. If $\lambda_{k} < 0$, gravity waves are
evanescent where $\omega^2 < N^2$. Instead, these waves propagate as
inertial waves in regions where $\omega^2 > N^2$. We do not examine these
inertial waves in detail because they do not satisfy the assumptions of the
traditional approximation.
However, the $m=-2$, $k=-2$ waves with
$\lambda_{k} > 0$ behave like gravity waves where $\omega^2 \ll
N^2$. We refer to them as r-waves. The r-waves satisfy the assumptions of the
traditional approximation and can be accurately characterized by our
techniques.

%%%%%%%%%%%%%%%%%%%%%%%%%%%%%%%%%%%%%%
\section{Tidal Response in Rotating White Dwarfs}
\label{tidal}

\subsection{Response to Tidal Forcing}

The extent to which an angular mode contributes to tidal processes is
largely determined by its overlap with the tidal potential, which can
be expanded in spherical harmonics. In the rotating frame of the
stars, the largest time-variable components are the $l=|m|=2$
components:
\be
\label{U}
U_{2,m}({\bf r},t) + U_{2,-m}({\bf r},t) = U(r) \Big[ Y_{2m}(\theta,\phi) e^{-i \omega t} + Y_{2m}^*(\theta,\phi) e^{i \omega t} \Big]
\ee
with  
\be
\label{U2}
U(r) = - \frac{G M' W_{22}}{a^3} r^2
\ee
and
\be
\label{om}
\omega = m (\Omega-\Omega_s),
\ee
with $\Omega$ the orbital frequency of the binary, $a$ the semi-major axis, and $W_{22}=\sqrt{3\pi/10}$. In the preceding (and following) sections we have chosen to work with the $e^{-i \omega t}$ term with $\omega>0$. Therefore when $\Omega > \Omega_s$, we use $m=2$ such that the orbit is prograde, and when $\Omega < \Omega_s$ we use $m=-2$ such that orbit is retrograde (in the rotating frame of the star).

The tidal potential appears as a forcing term in equations
\ref{xir1}-\ref{xip1}, but as noted above (see also Paper II), we
must subtract out the equilibrium tidal response to solve for the
dynamical tide. The forced oscillation equations for the dynamical part
of the tidal response are (see Appendix \ref{eqtide}) 
\be
\label{dp}
\frac{\partial}{\partial r} \delta P^{\rm dyn} = -\frac{g}{c_s^2} \delta P^{\rm dyn} + \rho \Big(\omega^2 - N^2\Big) \xi_r^{\rm dyn} + h_{klm} \rho \omega^2 \xi_r^{\rm eq} + h_{klm} m q \rho \omega^2 \xi_\perp^{\rm eq} ,
\ee
\be
\label{dxir}
\frac{\partial }{\partial r} \xi_r^{\rm dyn} = \bigg(\frac{g}{c_s^2} - \frac{2}{r}\bigg) \xi_r^{\rm dyn} + \bigg( \frac{\lambda_k}{\rho r^2 \omega^2} -  \frac{1}{\rho c_s^2} \bigg) \delta P^{\rm dyn} - \frac{h_{klm}l(l+1)}{r} \xi_\perp^{\rm eq}  - \frac{g_{klm}}{r} \xi_r^{\rm eq}.
\ee
These equations describe the forced response of a Hough wave
(specified by indices $k$ and $m$) due to one component of the tidal
potential (specified by indices $l$ and $m$). All quantities are
functions of $r$ only, as we have already integrated over their
angular dependence.

The key observation used to derive equations \ref{dp} and \ref{dxir}
is that the traditional approximation applies to the dynamical tide
{\it but not to the equilibrium tide}. Properly subtracting the
equilibrium tide leads to the appearance of the last terms in
equations \ref{dp} and \ref{dxir}, which are additional forcing terms
due to the Coriolis force on the equilibrium tide. These terms vanish
in the non-rotating ($q \rightarrow 0$) limit but are very important
when $q \gg 1$, and to our knowledge have not been taken into account
in any previous studies.

\begin{figure*}
\begin{centering}
\includegraphics[scale=.4]{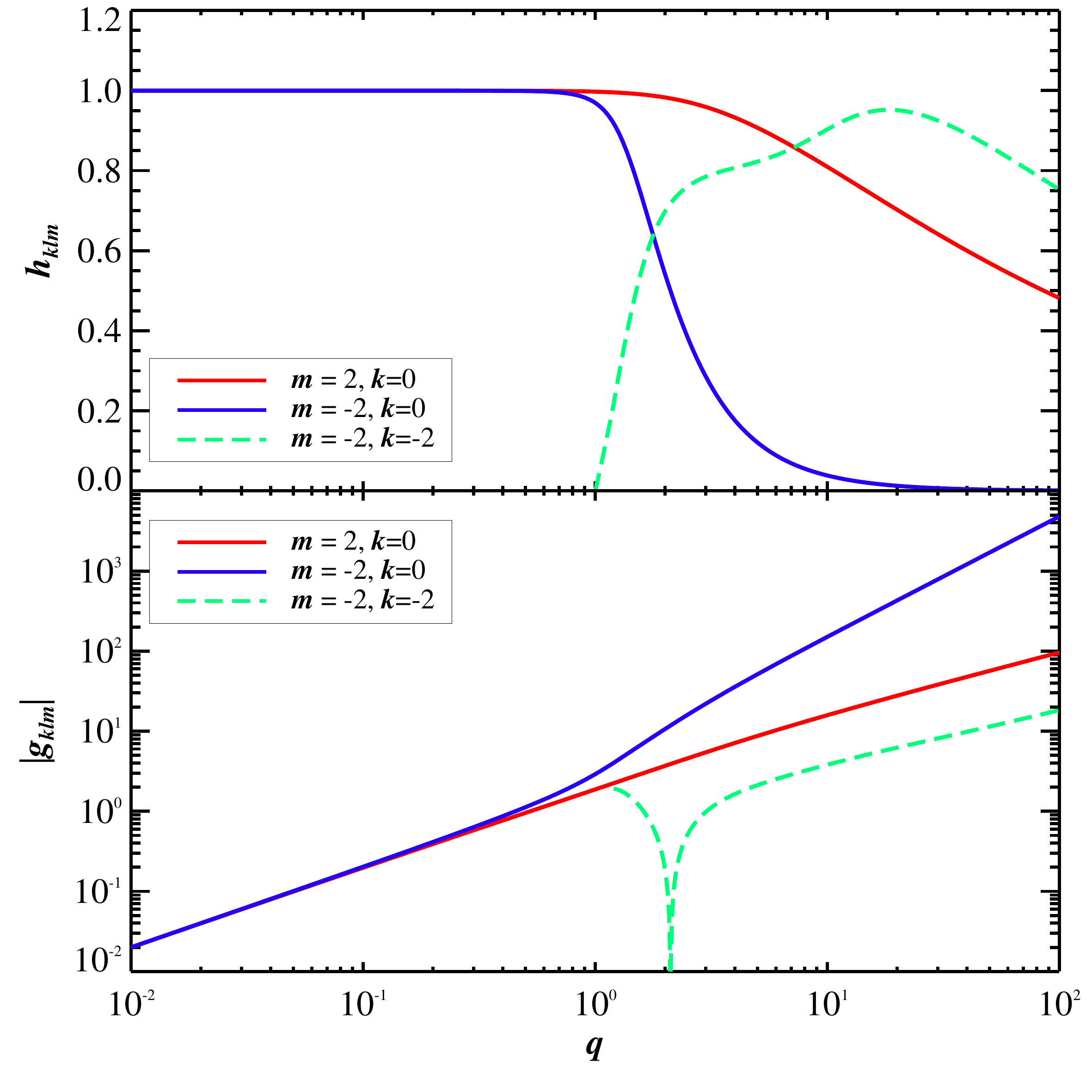}
\caption{\label{wdrotqh} Angular overlap integrals $h_{klm}$ and $|g_{klm}|$ (defined in equations \ref{hklm} and \ref{gklm}) with the $l=2$ component of the tidal potential as a function of $q$. The $k=-2$ r-wave solution does not exist for $q < 1$. }
\end{centering}
\end{figure*}

The coefficients $h_{klm}$ and $g_{klm}$ are angular overlap integrals
between Hough functions and spherical harmonics, and are given by
equations \ref{hklm} and \ref{gklm} in Appendix \ref{eqtide}. Figure
\ref{wdrotqh} shows values of $h_{klm}$ and $g_{klm}$ as a function of
$q$ for the $l=|m|=2$ component of the tidal potential. In the
non-rotating limit $q \rightarrow 0$, the Hough functions reduce to
spherical harmonics, $h_{klm} \rightarrow \delta_{k,l-m}$, and
$g_{klm} \rightarrow 0$. For $k=0$ waves, the value of $h_{k2m}$ falls
off monotonically with increasing $q$. For the prograde $m=2$ waves,
$h_{k2m}$ falls off slowly, and the $k=0$, $m=2$ gravity waves
dominate the tidal energy transfer at all values of $q$ for
subsyncronous rotation ($\Omega > \Omega_s$). However, $h_{k2m}$ falls
off much more rapidly for the $m=-2$ waves, diminishing their impact
on tidal energy transfer. Instead, the $k=-2$ waves become important,
with the value of $h_{k2m}$ approaching unity for $q \approx 20$, and
slowly decreasing for larger values of $q$. Therefore, while the
$k=0$, $m=-2$ gravity waves are most important at low values of $q$
for supersynchronous rotation ($\Omega < \Omega_s$), the $k=-2$,
$m=-2$ r-waves become important for $q \gtrsim 6$.\footnote{For $1
  \lesssim q \lesssim 6$, the traditional approximation does not
  accurately characterize the $k=-2$, $m=-2$ r-waves. It remains
  unclear whether the $k=-2$ branch of waves can contribute
  significantly to tidal dissipation in this range of $q$.}

For $q \gg 1$, we find $g_{k2m} \gg1$ for all branches of Hough waves
considered here. From Figure \ref{wdrotqh}, it is evident that when $q
\gg 1$, $g_{klm} \propto q$ for the $k=0$, $m=2$ branch and the
$k=-2$, $m=-2$ branch, whereas $g_{klm} \propto q^2$ for the $k=0$,
$m=-2$ branch (although we shall see below that this branch
contributes negligibly to tidal torque in WDs when $q \gg 1$). Thus,
the $g_{k2m}$ term can dominate the tidal forcing for large values of
$q$, and the failure to include it may lead to large errors in the
calculation of the dynamical tidal response.

\subsection{Angular Momentum Transport}

The tidally excited waves carry energy and angular momentum from the
region of excitation to the region of wave damping. As in Papers
II-IV, we assume the waves propagate into the outer layers of the
star, depositing their energy and angular momentum in the low density
subsurface regions where the waves become highly non-linear and break
(we justify this assumption in Section \ref{torque}, but see Papers II
and IV for more discussion on non-linear wave breaking). Consequently
we impose a radiative outer boundary condition\footnote{In our WD
  models we put out outer boundary at $r = 0.99 R$, below the surface
  convection zone, although the precise location is not important as
  long as it is above the radius at which the waves are excited, and
  in a stably-stratified region where the waves roughly obey WKB
  scaling relations.} as described in Paper II. The net tidal torque
on the star is then calculated from the angular momentum luminosity
flowing through the outer boundary.

The $z$-component of the wave angular momentum flux through a
spherical surface at radius $r$ is
\begin{align}
\label{angmom}
\dot{J}_z(r) &= \bigg\langle r^2 \int dA  \xi_r(r,\theta,\phi,t) \frac{\partial}{\partial \phi} \bigg(\delta P(r,\theta,\phi,t) + \rho U(r,\theta,\phi,t) \bigg) \bigg\rangle + cc \nonumber \\
		  &=  2 r^2 \int dA {\rm Re} \bigg[ \xi_r^*(r,\theta,\phi) \frac{\partial}{\partial \phi} \bigg(\delta P(r,\theta,\phi) + \rho U(r,\theta,\phi) \bigg) \bigg] \nonumber \\ 
		  &= 2 m r^2 \int dA {\rm Re} \bigg[ i \xi_r^{\rm dyn^*}(r,\theta,\phi) \delta P^{\rm dyn}(r,\theta,\phi) \bigg] \nonumber \\
		  &= 2 m r^2 {\rm Re} \bigg[ i \xi_r^{\rm dyn^*}(r) \delta P^{\rm dyn}(r) \bigg] \nonumber \\
		  &= 2 m \omega^2 \rho r^3 {\rm Re} \bigg[ i \xi_r^{\rm dyn^*}(r) \xi_\perp^{\rm dyn}(r) \bigg].
\end{align}
The first line in equation \ref{angmom} is derived in Appendix
\ref{angmomapp}, and is correct to second order in the wave
amplitude ($cc$ is the complex conjugate of the first term). We have emphasized the coordinate dependence of the
perturbation variables, as in equation \ref{xirang}, for clarity. The
second line follows from averaging the first line (the angle brackets
denote a time average), and the factor of two accounts for both the
wave displacement $\bxi$ and its complex conjugate. The third line
reduces the perturbations to their dynamical tide components, which we
justify in Appendix \ref{angmomapp}. The fourth line follows from
integration over the spherical surface, and utilizes the normalization
of Hough functions from equation \ref{norm}. Finally, the fifth line
uses the relation $\xi_\perp^{\rm dyn} \equiv \delta P^{\rm dyn}/(\rho r \omega^2)$ and shows the similarity between equation
\ref{angmom} and the familiar Eulerian result for non-rotating stars
given by equation 39 of Paper II.

\subsection{Stokes Drift Currents}
\label{stokesdrift}

We note that in a rotating star, the wave energy and
angular momentum fluxes cannot be simply derived from the Eulerian
perturbations as they are in Paper II and many other works (e.g., Lee
\& Saio 1993). The reasons can be traced back to the differences
between Eulerian and Lagrangian averaging discussed in Andrews \&
McIntyre (1978a,1978b), and are linked with the Stokes drift terms that enter at second order in the wave amplitude. 
Indeed, one can naively calculate the mass flux through a spherical surface using Eulerian perturbations,
\begin{align}
\label{mdot}
\dot{M}(r) &= \bigg\langle r^2 \int dA  \delta v_r(r,\theta,\phi,t) \delta \rho (r,\theta,\phi,t) \bigg) \bigg\rangle + cc \nonumber \\
		  &= 2 \omega r^2 {\rm Re} \bigg[ i \xi_r^*(r) \delta \rho (r) \bigg].
\end{align}
which in general is non-zero, reflecting the mass transported through
the surface by the Stokes drift. 

One can calculate the correct second-order wave flux by subtracting
out the contribution from Stokes drift. In the rotating frame, the Stokes drift velocity can be calculated from the linear wave solution via (Andrews \& McIntyre 1978a)
\be
\label{stokes}
{\bf v}^S = \overline{ (\bxi^* \cdot \bnab) \delta {\bf v} } + cc,
\ee
where the overbar indicates an average over azimuth and time. The Stokes drift adds contributions to the wave fluxes which are implicitly included in second-order expressions such as equation \ref{mdot}. One can explicity calculate the Stokes drift contribution
from the Stokes drift velocity. For instance, the Stokes drift contribution to the mass flux is
\be
\label{stokesm}
\dot{M}^S(r) = r^2 \int dA \rho v^S_r.
\ee
Some algebra shows that the mass flux of equation \ref{mdot} is due entirely due to the Stokes contribution of equation \ref{stokesm}. To calculate the ``wave" contribution we simply subtract the Stokes drift contribution, therefore the mass flux due to the waves is $\dot{M}(r)=0$ as we should expect. 

We can perform a similar procedure to calculate the wave angular momentum flux from the Reynold's stress, as long as we include the torque exerted by the Coriolis force on the Stokes drift current. The end result is that the wave angular momentum flux is given by equation \ref{angmom}. We also note that equation \ref{angmom} implies (see Appendix \ref{angmomapp}) that in the rotating frame,
\be
\label{enang}
\dot{J}(r) = \frac{m}{\omega} \dot{E}(r).
\ee
This relation holds regardless of the rotation rate or the latitudinal dependence of the fluid perturbations, in accordance with Kumar et al.~(1999).

Stokes currents are a real physical effect and are generated by
propagating waves in the star. The currents cause fluid elements to
drift over time, allowing them to transport mass, energy, angular
momentum, metals, etc. While equation (\ref{angmom}) accurately
calculates the net tidal torque on the star, the Stokes currents may
be important for the redistribution of angular momentum within the
star. Additionally, a counter-current must be produced (akin to the
undertow produced by breaking waves at a beach) to balance the mass
transport via Stokes currents. These counter-currents are not captured
by linear calculations, and they need not be equal and opposite to the
Stokes drift. The Stokes drift current and counter-current may
therefore set-up a circulation which could enhance mixing processes,
and which could generate or reduce differential rotation. Quantifying
these effects is beyond the scope of this paper, but they should be
investigated in future studies.

\section{Tidal Torque: Numerical Results}
\label{torque}

We have now assembled all the ingredients necessary to calculate the
tidal torque on 
white dwarfs in compact binaries. Given an orbital
frequency $\Omega$ and spin frequency $\Omega_s$, the $l=|m|=2$
component of the tidal potential has tidal forcing frequency
$\omega=|2(\Omega-\Omega_s)|$ and spin parameter
$q=2\Omega_s/\omega$. We first calculate the Hough functions $H_{k}(q)$
for the branches $k$ that contribute to the tidal torque, and the
associated overlap integrals $h_{klm}$ and $g_{klm}$. We then 
solve equations \ref{dp} and \ref{dxir} with the outgoing wave outer
boundary condition, as discussed in Paper II, and use equation
\ref{angmom} to calculate the associated tidal torque. In appendix
\ref{innerbc}, we discuss the inner boundary condition and subtleties
that can arise when using the traditional approximation.

It is useful to express the tidal torque in dimensionless units (see
Papers II and III) such that
\be
\label{Jdot}
T_{\rm tide} = \dot{J}_z(r_{\rm out}) = T_0 F(\omega,q),
\ee
with
\be
\label{T0}
T_0 = \frac{G M'^2}{a} \bigg(\frac{R}{a}\bigg)^5.
\ee
The quantity $F(\omega,q)$ is the dimensionless tidal torque and is
related to the parameterized tidal dissipation factor $Q$ (Goldreich
\& Soter 1966; Alexander 1973; Hut 1981) by 
$|F(\omega,q)|=3k_2/(2Q)$, where $k_2$ is the tidal Love number.
Note that $F>0$ for $\Omega>\Omega_s$ and $F<0$ for $\Omega<\Omega_s$.

\begin{figure*}
\begin{centering}
\includegraphics[scale=.5]{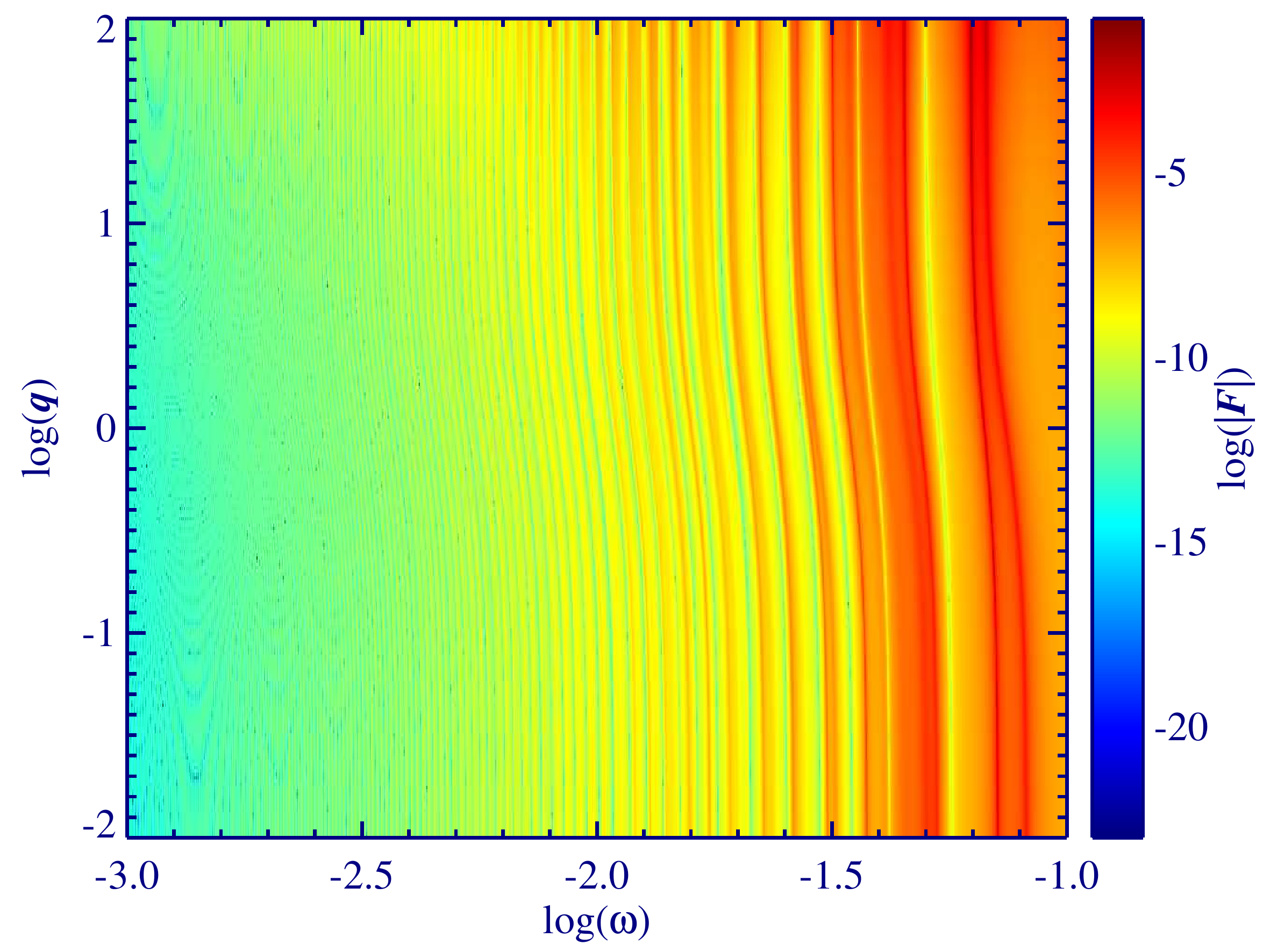}
\caption{\label{wdrotl2m2} Dimensionless tidal torque $F(\omega,q)$ as a function of tidal forcing frequency $\omega = m (\Omega-\Omega_s)$ and rotation parameter $q = 2 \Omega_s/\omega$, for the $M = 0.6 M_\odot$, $T_{\rm eff} = 10^4 {\rm K}$ WD model described in Paper II. This plot shows the torque exerted by the excitation of $m=2$, $k=0$ prograde g-waves responsible for tidal dissipation in sub-synchronously rotating WDs. In the limit $q \rightarrow 0$, this plot is identical to the results of Figure 7 of Paper II.}
\end{centering}
\end{figure*}

\begin{figure*}
\begin{centering}
\includegraphics[scale=.5]{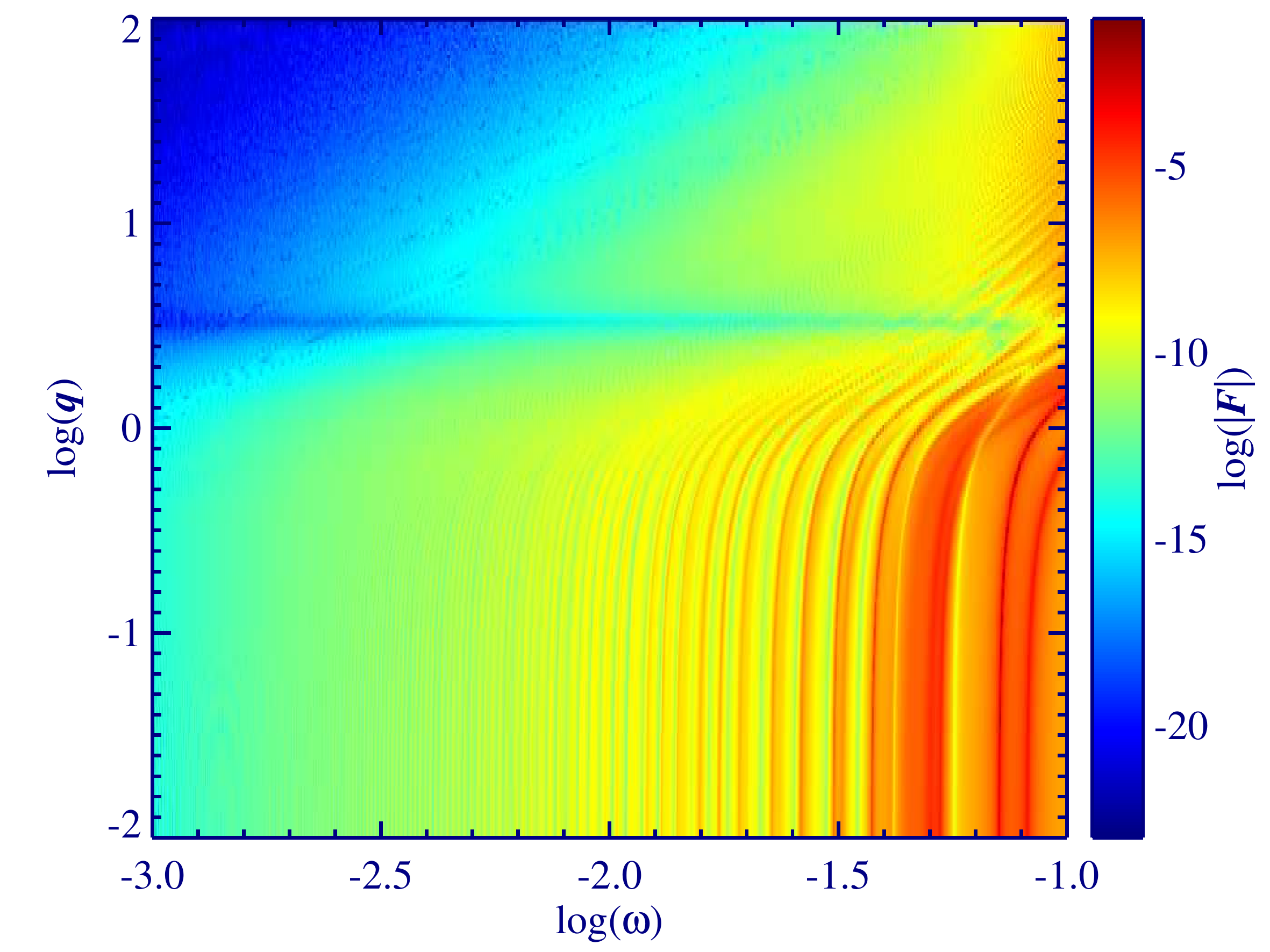}
\caption{\label{wdrotl2m2n} Same as Figure \ref{wdrotl2m2}, but for $m=-2$, $k=0$ retrograde g-waves excited in super-synchronously rotating WDs.
Note that $F$ is negative in this case.}
\end{centering}
\end{figure*}

\begin{figure*}
\begin{centering}
\includegraphics[scale=.5]{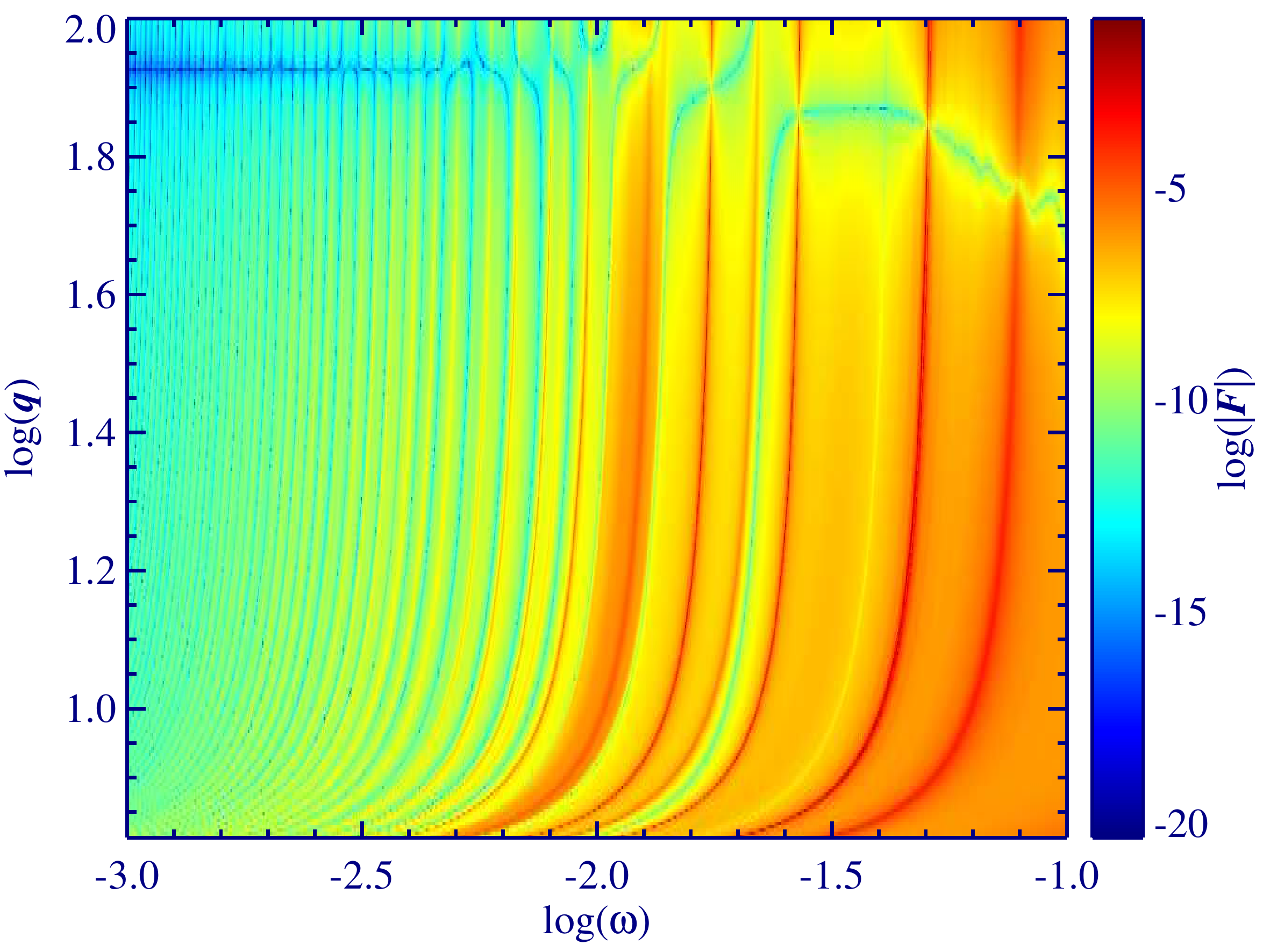}
\caption{\label{wdrotl2m2r} Same as Figure \ref{wdrotl2m2}, but for $m=-2$, $k=-2$ retrograde r-waves excited in super-synchronously rotating WDs. We have only plotted our results for $q > 6$ because the traditional approximation does not accurately characterize these r-waves for $q < 6$.
Note that $F$ is negative in this case.}
\end{centering}
\end{figure*}

Figures \ref{wdrotl2m2}-\ref{wdrotl2m2r} show our calculations of
$F(\omega,q)$ for $m=2$ and $k=0$ (prograde g-waves), $m=-2$ and $k=0$
(retrograde g-waves), and $m=-2$ and $k=-2$ (retrograde r-waves),
respectively. 
Recall that our convention for the perturbation is $e^{im\phi-i\omega t}$
with $\omega>0$, so $m=-2$ corresponds to $\Omega<\Omega_s$, for which $F<0$.
These calculations are based on the same $M=0.6
M_\odot$, $T_{\rm eff} = 10^4 K$, CO WD model considered in Paper II,
and the results in Figure \ref{wdrotl2m2} for $q \simeq 0$ are
identical to those of Paper II. The nearly vertical ridges evident in
Figures \ref{wdrotl2m2}-\ref{wdrotl2m2r} reveal the oscillatory
dependence of $F(\omega,q)$ on the forcing frequency, while the
increase in $F(\omega,q)$ towards larger values of $\omega$ reflects
the approximate scaling $F(\omega,q) \propto \omega^5$ scaling found
in Paper II. In Figures \ref{wdrotl2m2} and \ref{wdrotl2m2n}, the bend
in the vertical structure at $q \approx 1$ is caused primarily by the
changing eigenvalue $\lambda_k$. Since the radial wavenumber 
for g-waves scales as
$k_r^2 \simeq \lambda_k N^2/\omega^2$, the changing value of
$\lambda_k$ affects the radial wavelength of the tidally excited waves
and thereby alters the \textquotedblleft resonant" forcing frequencies
at which the tidal response is maximized (see Paper II).

Figure \ref{wdrotl2m2} shows the tidal response for the prograde
g-waves that dominate tidal spin-up for sub-synchronous WDs ($\Omega>\Omega_s$).
Other than the slight bend in in the resonant ridges at $q \approx 1$,
the value of $F(\omega,q)$ is not strongly dependent on the value of
$q$ and hence is relatively insensitive to the rotation of the WD (as
long as the WD spins sub-synchronously). Consequently, the tidal
torque and heating rates derived in Papers II and IV are not strongly
modified by the presence of the Coriolis force, and the general
conclusions of those works remain valid.

The tidal response for retrograde waves in super-synchronous WDs ($\Omega<\Omega_s$), 
shown in Figures \ref{wdrotl2m2n} and \ref{wdrotl2m2r}, exhibits
considerably more dependence on the value of $q$. The retrograde
g-waves (with $k=0$ and $m=-2$) are strongly modified by the Coriolis
force, primarily because their angular eigenvalue $\lambda_k$
increases rapidly for $q \gtrsim 1$ (for these waves $\lambda_k
\propto q^2$ for $q \gg 1$, Bildsten \& Cutler 1996). Not only does
this cause a strong bend in the location of the resonant ridges, 
it also strongly reduces the tidal torque exerted by these waves. Since the
value of $\lambda_k$ increases with $q$, the radial wavelength of
these waves decreases, and their overlap with the tidal potential is
diminished. From equation 69 of Paper II, we find an approximate
scaling $F(\omega,q) \propto \lambda_k^{-5/2} \propto q^{-5}$ for $q
\gg 1$. Therefore the retrograde g-waves provide a miniscule tidal
torque at large values of $q$.

However, as the importance of the retrograde g-waves falls off for $q
\gtrsim 1$, the $k=-2$, $m=-2$ branch of r-waves replace them and
become the dominant form of tidal dissipation. Figure \ref{wdrotl2m2r}
shows that the r-waves create a tidal response analagous to the
pro-grade g-waves, with similar resonant ridges, and similar values of
$F(\omega,q)$ (although note the different plot range in Figure
\ref{wdrotl2m2r}, as we have only plotted the region with $q>6$ where
the eigenvalue $\lambda_k$ is positive). This is not too surprising,
since we showed in Section \ref{Hough} that the $k=-2$ r-waves behave
like g-waves for large values of $q$.

\begin{figure*}
\begin{centering}
\includegraphics[scale=.6]{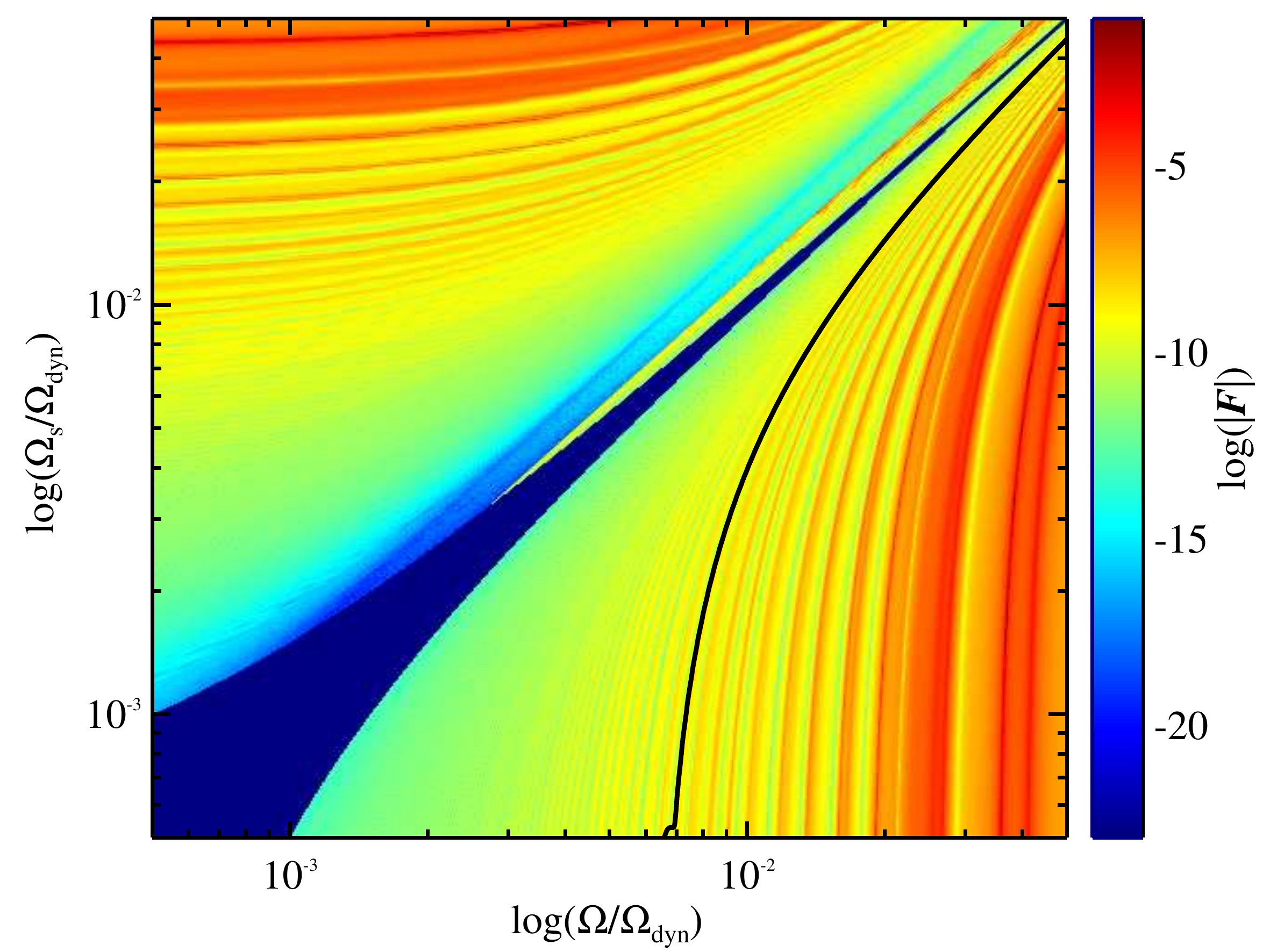}
\caption{\label{wdrot2} The dimensionless tidal torque $|F(\omega,q)|$ as a function of the orbital frequency $\Omega$ and spin frequency $\Omega_s$. This figure shows the combined torques from Figures \ref{wdrotl2m2}-\ref{wdrotl2m2r}. The dark blue region has $\omega < 10^{-3} \Omega_{\rm dyn}$ and has not been calculated. The thick black line shows the orbital and spin evolution of an inspiraling WD binary as described in the text. }
\end{centering}
\end{figure*}

Figure \ref{wdrot2} shows the dimensionless tidal torque $F(\omega,q)$
as a function of dimensionless orbital and spin frequency
($\bar{\Omega} = \Omega/\Omega_{\rm dyn}$, $\bar{\Omega}_s =
\Omega_s/\Omega_{\rm dyn}$).  The diagonal dark blue slice through the
middle of the plot is the locus of points where spin and orbit are
nearly synchronized such that $\omega < 10^{-3} \Omega_{\rm dyn}$, and
lies outside of the range of our calculations. The top-left half of
the plot represents super-synchronous rotation where retrograde waves
cause tidal dissipation, while the bottom right represents
sub-synchronous rotation where prograde waves operate. The r-waves
operate in the narrow slice just above the diagonal. It is evident
from this plot that tidal dissipation is not symmetric across the line
$\Omega=\Omega_s$, as it would be in the absence of the Coriolis
force. In the r-wave region, tidal dissipation is generally stronger
than it is for the same value of $|\omega|$ corresponding to prograde
g-waves, primarily because the r-waves have smaller values of
$\lambda_k$ and thus longer wave lengths, making them couple more effectively
to the tidal potential. Conversely, in the retrograde g-wave region
where $q\approx4$ (which appears as the blue slice above the r-wave
region), neither r-waves nor g-waves are effective in generating tidal
dissipation (although a precise understanding of the nature of tidal
dissipation in this regions remains hampered by our use of the
traditional approximation).

\begin{figure*}
\begin{centering}
\includegraphics[scale=.6]{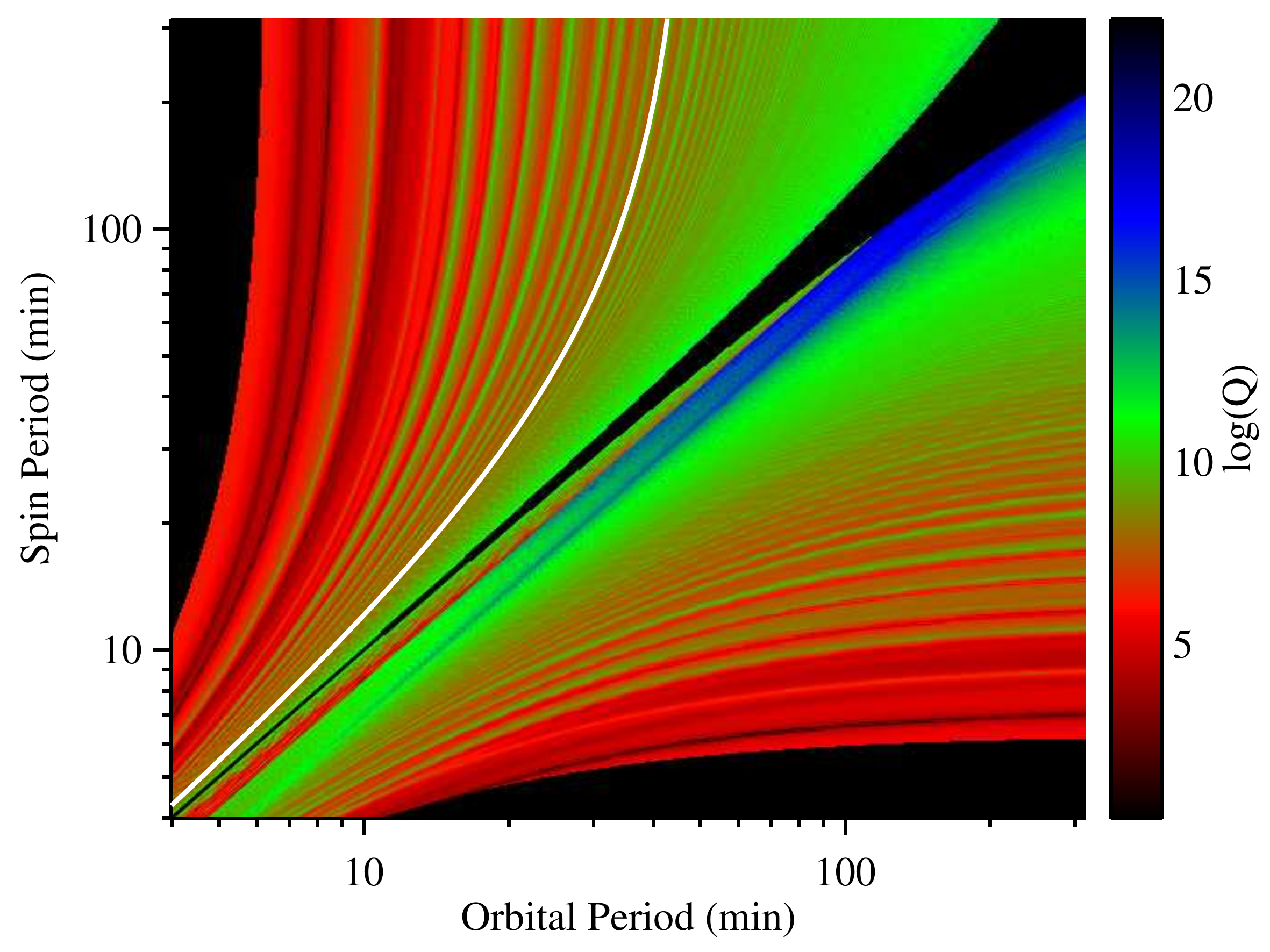}
\caption{\label{wdrot3} Tidal $Q$ parameter of the $0.6 M_\odot$ WD model used in Figures \ref{wdrotl2m2}-\ref{wdrot2}, as a function of its orbital period and spin period. The black regions have $\omega > 10^{-1} \Omega_{\rm dyn}$ or $\omega < 10^{-3} \Omega_{\rm dyn}$ and have not been calculated. The thick white line shows the orbital and spin evolution of a WD binary as described in the text.}
\end{centering}
\end{figure*}

Finally, we plot the tidal torque in terms of an effective tidal $Q$
in Figure \ref{wdrot3}, as a function of the WD orbital period $P$ and
spin period $P_s$. In a similar fashion to Figure \ref{wdrot2}, the
prograde g-waves occupy the upper left part of the figure, the
retrograde g-waves occupy the lower right part, and the retrograde
r-waves occupy the sliver just below the synchronization slice ($P
\approx P_s$). We can see that the effective tidal $Q$ varies by many
orders of magnitude in the parameter space investigated, and choosing
a single value of $Q$ to apply to tidal interactions in WDs is a naive
procedure. Nonetheless, as discussed in Papers II and IV, the combined
spin and orbital evolution of a compact WD binary may lead to a
scenario in which the tidal forcing frequency $\omega$ and effective
value of $Q$ remain nearly constant as the WD orbit decays and the WD
is spun up to near synchronization.

We therefore perform orbital evolution calculations similar to those
of Papers II and IV, only with $F(\omega,q)$ now
dependent on both $\omega$ and $q$. The results are shown by the black
and white lines in Figures \ref{wdrot2} and \ref{wdrot3}, for an equal
mass ($M=0.6 M_\odot$) WD binary with no initial spin ($\Omega_s=0$)
and initial orbital period of $P=5 {\rm hr}$. We find that the g-waves
are indeed strongly non-linear ($\xi_r k_r > 1$) at the short orbital
periods shown in Figure \ref{wdrot3}, validating the use of the
radiative outer boundary condition.\footnote{At longer periods, ($P
  \gtrsim 45 {\rm min}$), the waves are not always strongly
  non-linear, and may set up standing oscillation modes. In this case,
  the radiative outer boundary condition is not appropriate, and one
  must analyze g-mode excitation as discussed in Paper I and Burkart
  et al. 2013.} In a manner nearly identical to that of non-rotating
WDs, the combined effects of gravitational wave-induced orbital decay
and tidal spin-up act to lock the system into a state of nearly
constant $\omega$ and $Q$ at short orbital periods. Thus, the
evolutionary tracks in Figures \ref{wdrot2} and \ref{wdrot3} trace out
nearly constant values of $\omega$, with a corresponding value of $Q
\approx 10^7$.

\section{Discussion and Conclusions}
\label{conc}

We have examined the effect of the Coriolis force on the excitation of
dynamical tides in rotating stars, with specific application to white
dwarfs in compact binary systems. In particular, we study the
dependence of the tidal dissipation on the tidal forcing frequency
$\omega=m(\Omega-\Omega_s)$ and rotational parameter
$q=2\Omega_s/\omega$. We utilize the traditional approximation to
examine the effect of the Coriolis force, which applies well to the
gravito-inertial waves important for tidal dissipation in WDs. As one
might expect, the Coriolis force is negligible when $q \ll 1$ and
important when $q \gtrsim 1$.

There are four primary effects of the Coriolis force. First, it alters
the angular profile of g-waves propagating in stars, changing their
effective angular wavenumber $\lambda_k$ and their angular overlap
with the tidal potential. Second, the changing value of $\lambda_k$
also changes the radial wavelength of the waves, modifying their
radial overlap with the tidal potential. Third, the Coriolis force
introduces new classes of waves (inertial waves and r-waves) which
rotationally mix with g-waves and provide strong tidal dissipation in
certain regions of parameter space. Fourth, 
the coupling of the equilibrium tidal distortion with the Coriolis force
introduces new tidal forcing terms, which dominate the tidal response of the star
when $q \gg 1$.

We find that the Coriolis force has 
a small qualitative effect on tidal
dynamics in sub-synchronous WDs in which prograde g-waves create most
of the tidal torque. For these waves, the value of $\lambda_k$ is not
very dependent on $q$, and thus the angular and radial structure of
the waves is not strongly affected. Moreover, in the traditional
approximation, no prograde r-wave branches obtain positive values of
$\lambda_k$, and thus there are no additional types of prograde waves
that can propagate through the stably stratified interiors of WDs and
cause tidal dissipation.

By contrast, the Coriolis force has a large effect on the tidal
dynamics of super-synchronous WDs in which retrograde g-waves and
r-waves dominate the tidal torque. The retrograde $k=0$, $m=-2$ g-waves
(which correspond to $l=2$, $m=-2$ waves for a slowly spinning star)
that dominate tidal dissipation at small values of $q$ are strongly
altered by the Coriolis force. Their angular wave number increases as
$\lambda_k \propto q^2$, causing their radial wavelength to decrease,
strongly reducing their overlap with the tidal potential and making
them irrelevant for tidal dissipation when $q \gg 1$. However, the
Coriolis force also introduces an r-wave branch of solutions for $q>1$,
which obtains positive $\lambda_k$ for $q > 6$ (for the $k=-2$
r-waves, under the traditional approximation), leading to g-wave-like
behavior. The r-waves maintain a small but positive value of
$\lambda_k$ for $q \gg 1$, allowing them to couple strongly with the
tidal potential and dominate tidal dissipation in this region of
parameter space.

The above mentioned waves have also been identified in previous
works. In the language of Pantillon et al. (2007) and Mathis et
al. (2008), the prograde ($m=2$, $k=0$) waves are the $s=-1$ Kelvin
waves (note the opposite definition of the sign of $m$ in those
works). The retrograde g-waves ($m=-2$, $k=0$) are $s=1$
gravito-inertial waves, and the retrograde r-waves ($m=-2$, $k=-2$)
are $s=1$ Rossby waves. The Yanai waves discussed in the above works
are anti-symmetric across the equator and thus not relevant to our
problem, but may be relevant for tidal dissipation when the spin and
orbital axes are misaligned.

It is also important to address the limitations of the traditional
approximation, namely, that it does not accurately characterize the
oscillation mode spectrum of a star for $q>1$, as pointed out by
Mathis et al. (2008) and examined in detail by Dintrans et al. (1999) and
Dintrans \& Rieutord (2000).\footnote{Our solutions are analagous to the
  $E_1$ modes of Dintrans et al. 1999, for $q < 1$, and to the $H_2$
  modes for $q>1$. Similar to Figure 1c of Dintrans et al. 1999,
  g-waves in the traditional approximation are confined to the
  equatorial regions of the star. It may be possible that modes
  similar to the $H_1$ and $E_2$ modes of Dintrans et al. 1999 can
  exist near the core of the star where $N\sim 2 \Omega_s$.} For
$q>1$, the oscillation equations become of a mixed hyperbolic and
elliptic type, with the solutions exhibiting wave attractors such that
the waves are focused onto rays where dissipation becomes very
large. We have partially circumvented this problem because we consider
outgoing waves which are not reflected from an outer boundary, and
which cannot exhibit the periodic orbits and singularities discussed
in Dintrans et al. (1999). Since the solutions we consider for $q>1$ do
not directly violate any of the assumptions of the traditional
approximation (this applies to the r-wave branch so long as we
consider regions with a positive eigenvalue $\lambda_k$), it appears
that our results remain qualitatively and quantitatively
accurate. However, we cannot rule out the possibility that additional
forms of dissipation exist, e.g., due to dynamical tides composed of
waves which are able to reflect near the surface of the WD and are
focused onto wave attractors.

We have used our results to calculate the orbital and spin evolution of
WDs in tight binary systems. Since the Coriolis force does not
strongly modify the prograde g-waves that induce tidal dissipation in
sub-synchronously rotating inspiraling WDs, the general conclusions of
Papers II-IV and Burkart et al. (2013) remain valid. Thus, compact WDs
whose orbits decay via gravitational radiation should be spun up
nearly to synchronization by the time Roche-Lobe overflow occurs, with
the tidal forcing frequency $\omega$ and effective tidal $Q$ remaining
nearly constant for orbital periods less than approximately 1 hour.

We can also speculate on the tidal dynamics of a WD accreting material
from a companion such that the WD begins to spin
super-synchronously. Our analysis indicates that the retrograde r-wave
branch is capable of producing strong tidal dissipation as the WD is
spun up and the tidal forcing frequency $|\omega|$ increases. Thus, we
do not expect WDs accreting in compact systems (e.g., AM CVn systems)
to be spun up to near break up, as tidal dissipation will likely
transfer some of the accreted angular momentum back to the orbit of
the donor star.\footnote{This statement assumes the accreted angular
  momentum can be quickly transferred from the accreted layers into
  the core of the WD, although the precise physical mechanism and
  timescale for this angular momentum transport remain unknown.} The
spin evolution of an accreting WD in a CV system is less certain
because the companion is more distant and the g-waves/r-waves may
remain linear, in which case 
the excitation of discrete g-modes studied in Paper I and Burkart et al. (2013)
become important (although these discrete g-modes must be modified
to account for rotational effects). Whether tides can act fast enough
to stabilize the mass transfer from a degenerate companion (as
suggested in Nelemans et al. 2001b and Marsh et al. 2004) remains
unclear, as this scenario hinges on a delicate balance of several
physical mechanisms (e.g., the gravitational-wave-induced orbital
decay, the mass-transfer induced orbital expansion, the tidal torque
on the orbit, and the response of the donor star to mass loss). We
plan to investigate this issue in a future paper.

\section*{Acknowledgments}
We thank Peter Goldreich and Rich Townsend for useful discussions. J. Fuller
acknowledges partial support from NSF under grant No. AST-1205732 and
through a Lee DuBridge Fellowship at Caltech. This work has been
supported in part by NSF grants AST-1008245, 1211061, PHY11-25915, and NASA grants
NNX12AF85G and NNX10AP19G.

\appendix

\section{Forced Oscillation Equations for Dynamical Tides in Rotating Stars}
\label{eqtide}

In this appendix we derive equations \ref{dp} and \ref{dxir} for the
dynamical part of the tidal response of a rotating star. Although the
dynamical part of the forced oscillation equations has been derived in
previous studies (e.g., Paper II), the result changes considerably in
a rotating star in which the latitudinal dependence of the waves is
altered. The key observation is that while the equilibrium tide has a
spherical harmonic angular dependence, the dynamical tide has a Hough
function angular dependence. We must therefore be careful in our
separation of these tidal components. We begin with the momentum
equations (\ref{xir1})-(\ref{xip1}):
\be
\label{xir1a}
-\rho \omega^2 \xi_r = -\frac{\partial}{\partial r} \Big( \delta P + \rho U \Big) + U \frac{\partial \rho}{\partial r} - \frac{g}{c_s^2} \delta P - \rho N^2 \xi_r - 2i\Omega_s \omega \rho \xi_\phi \sin\theta,
\ee
\be
\label{xit1a}
-\rho \omega^2 \xi_\theta = -\frac{1}{r} \frac{\partial}{\partial \theta} \Big( \delta P + \rho U \Big) - 2i\Omega_s \omega \rho \xi_\phi \cos\theta,
\ee
\be
\label{xip1a}
-\rho \omega^2 \xi_\phi = -\frac{1}{r \sin \theta} \frac{\partial}{\partial \phi} \Big( \delta P + \rho U \Big) + 2i\Omega_s \omega \rho \big(\xi_\theta \cos \theta + \xi_r \sin \theta \big),
\ee
the adiabatic relation 
\be
\label{drhoa}
\delta \rho = \frac{1}{c_s^2} \delta P + \frac{\rho N^2}{g} \xi_r,
\ee
and the continuity equation
\be
\label{conta}
\delta \rho + \frac{1}{r^2} \frac{\partial}{\partial r} \big( \rho r^2 \xi_r \big) + \rho \bnab_\perp \cdot \bxi_\perp=0.
\ee
These equations apply for a tidally forced, spherical
(non-centrifugally distorted) star in the linear and adiabatic limits,
under the Cowling approximation.

The equilibrium tide is found by taking the limit $\omega \rightarrow 0$, yielding 
\be
\xi^{\rm eq}_r = - U/g,
\ee
\be
\bxi^{\rm eq}_\perp = -\frac{1}{l(l+1)} \bnab_\perp \frac{\partial}{\partial r} \bigg( \frac{U r^2}{g} \bigg) ,
\ee
\be
\delta P^{\rm eq} = - \rho U,
\ee
and here $l$ is the component of the tidal potential (given in
equations \ref{U} and \ref{U2}) we are considering. We restrict our
analysis to the $l=|m|=2$ case since these components dominate for
nearly circular, aligned orbits at the orbital periods of interest.

We now let $\bxi = \bxi^{\rm dyn} + \bxi^{\rm eq}$ and $\delta P =
\delta P^{\rm dyn} + \delta P^{\rm eq}$ and put these expressions into
equation \ref{xir1a}-\ref{conta}. We apply the traditional
approximation by ignoring the $\xi_\phi^{\rm dyn}$ term in equation
\ref{xir1a} and the $\xi_r^{\rm dyn}$ term in equation \ref{xip1a},
{\it but we do not drop the equilibrium tide components} since
they are not negligible. After some rearranging we find
\be
\label{xir2a}
\xi^{\rm dyn}_r \bigg(1 - \frac{N^2}{\omega^2} \bigg) = \frac{1}{\rho \omega^2} \bigg[\frac{\partial}{\partial r} + \frac{g}{c_s^2} \bigg] \delta P^{\rm dyn} - \xi_r^{\rm eq}(r)Y_{lm} - m q \xi_\perp^{\rm eq}(r) Y_{lm},
\ee
\be
\label{xit2a}
\xi_\theta^{\rm dyn} = \frac{1}{\rho r \omega^2} \frac{1}{1-q^2 \cos^2 \theta} \bigg[ \frac{\partial}{\partial \theta} - m q \frac{\cos \theta}{\sin \theta} \bigg] \delta P^{\rm dyn} - \xi_\perp^{\rm eq}(r) \frac{\partial}{\partial \theta} Y_{lm} + \frac{q^2 \sin \theta \cos \theta}{1-q^2 \cos^2 \theta} \xi_r^{\rm eq}(r) Y_{lm},
\ee
\be
\label{xip2a}
\xi_\phi^{\rm dyn} = \frac{1}{\rho r \omega^2} \frac{1}{1-q^2 \cos^2 \theta} \bigg[ \frac{i m}{\sin \theta} - i q \cos \theta \frac{\partial}{\partial \theta} \bigg] \delta P^{\rm dyn} - \frac{i m}{\sin \theta} \xi_\perp^{\rm eq}(r) Y_{lm} - \frac{i q \sin \theta}{1 - q^2 \cos^2 \theta} \xi_r^{\rm eq}(r) Y_{lm},
\ee
\be
\label{dP2a}
\delta P^{\rm dyn} = \rho g \xi_r^{\rm dyn} - \frac{\rho c_s^2}{r^2} \frac{\partial}{\partial r} \Big( r^2 \xi_r^{\rm dyn} \Big) - \rho c_s^2 \bnab_\perp \cdot \bxi_\perp^{\rm dyn}.
\ee
Here, we have explicitly expressed the angular dependence of the
equilibrium tide terms in terms of spherical harmonics, and defined
\be
\xi_\perp^{\rm eq}(r) = -\frac{1}{l(l+1)r} \frac{\partial}{\partial r} 
\bigg[\frac{U(r) r^2}{g} \bigg].
\ee

Equations \ref{xir2a}-\ref{xip2a} show that we can think of the
dynamical tide as being forced by the equilibrium tidal distortion
rather than by the tidal potential. Note that the last term in
equations \ref{xir2a}-\ref{xip2a} are additional tidal forcing terms
that vanish in the limit $q \rightarrow 0$. The spin frequency
$\Omega_s$ does not explicitly appear in equations
\ref{xir2a}-\ref{xip2a} because we have used $2 \Omega_s = q \omega$.

Substituting equations \ref{xit2a} and \ref{xip2a} into equation \ref{dP2a}, we obtain
\begin{align}
\label{dP3a}
\delta P^{\rm dyn} &= \rho g \xi_r^{\rm dyn} - \frac{\rho c_s^2}{r^2} \frac{\partial}{\partial r} \Big( r^2 \xi_r^{\rm dyn} \Big) - \rho c_s^2 \bigg[ \frac{1}{\rho r^2 \omega^2} \mathcal{L} \ \delta P^{\rm dyn} \ + \nonumber \\
 &\frac{l(l+1)}{r} \xi_\perp^{\rm eq} Y_{lm} - \frac{q^2}{r} \xi_r^{\rm eq} \frac{\partial}{\partial \mu} \bigg( \frac{1}{1 -q^2\mu^2} \mu (1-\mu^2) Y_{lm} \bigg) + \frac{qm}{r} \xi_r^{\rm eq} \frac{1}{1-q^2\mu^2} Y_{lm} \bigg].
\end{align}
Here, $\mathcal{L}$ is the Laplace tidal operator from equation \ref{lop}, and $\mu=\cos \theta$. We now decompose the dynamical tide into Hough functions $H_{k}$, and integrate equations \ref{xir2a} and \ref{dP3a} over a spherical surface to obtain 
\be
\label{xir3a}
\xi^{\rm dyn}_{r} \bigg(1 - \frac{N^2}{\omega^2} \bigg) = \frac{1}{\rho \omega^2} \bigg[\frac{\partial}{\partial r} + \frac{g}{c_s^2} \bigg] \delta P^{\rm dyn} - h_{klm} \Big( \xi_r^{\rm eq} + m q \xi_\perp^{\rm eq} \Big),
\ee
and
\begin{align}
\label{dP4a}
\delta P^{\rm dyn} \bigg(1 - \frac{\lambda_k c_s^2}{r^2 \omega^2} \bigg) &= \frac{\rho c_s^2}{r^2} \bigg[ \frac{g}{c_s^2} - \frac{\partial}{\partial r} \bigg] \Big( r^2 \xi_r^{\rm dyn} \Big) - \rho c_s^2 \bigg[ \frac{l(l+1)h_{klm}}{r} \xi_\perp^{\rm eq} + \frac{g_{klm}}{r} \xi_r^{\rm eq} \bigg].
\end{align}
The angular overlap integrals $h_{klm}$ and $g_{klm}$ describe the
angular coupling with the tidal potential and are given by
\be
\label{hklm}
h_{klm} = \int\! dA \, Y_{lm} H_{k} e^{-i m \phi}
\ee
and 
\begin{align}
\label{gklm}
g_{klm} &= q^2 \int dA \frac{1}{\sin \theta} \frac{\partial}{\partial \theta} \bigg( \frac{\sin^2 \theta \cos \theta}{1-q^2\cos^2 \theta} Y_{lm} \bigg) H_{k} e^{-i m \phi} 
+ q m \int dA \frac{1}{1-q^2 \cos^2 \theta} Y_{lm} H_{k} e^{-i m \phi} \nonumber \\
	&= -q^2 \int dA  \bigg( \frac{\sin \theta \cos \theta}{1-q^2\cos^2 \theta} Y_{lm} \bigg) \frac{\partial}{\partial \theta} H_{k} e^{-i m \phi} 
+ q m \int dA \frac{1}{1-q^2 \cos^2 \theta} Y_{lm} H_{k} e^{-i m \phi} \nonumber \\
	& =  i q \int dA \sin \theta \xi_r^{\rm eq}(\theta,\phi) \xi_\phi^{\rm dyn^*}(\theta,\phi),
\end{align}
with the second equality following from integrating the first term by
parts. In the third equality, we have used definitions $\xi_r^{\rm
  eq}(\theta,\phi)=Y_{lm}(\theta,\phi)$ and $\xi_\phi^{\rm
  dyn^*}(\theta,\phi)$ given by the angular part of the right hand
side of equation \ref{xipang} to highlight the origin of the angular
integral composing $g_{klm}$. Equations \ref{xir3a} and \ref{dP4a} are
identical to equations \ref{dp} and \ref{dxir} and can be numerically
integrated to find the dynamical component of the tidal response. The
last terms in equations \ref{xir3a} and \ref{dP4a} vanish for a
non-rotating star, and do not appear if one mistakenly applies the
traditional approximation to the equilibrium component of the tidal
response. For $q \gg 1$, these terms dominate the tidal forcing, and
so they must be included.

\section{Angular Momentum Flux in Rotating Stars}
\label{angmomapp}

\subsection{Angular Momentum Flux through a Surface}

In this appendix we provide a simple derivation of the angular
momentum flux through the quasi-spheriodal surface of a perturbed
spherical shell. The derivation is simplest if we adopt an inertial
frame of reference. We begin from the $z$-component of the angular
momentum equation:
\be
\label{am1}
\frac{\partial}{\partial t} j_z + \bnab \cdot \big( j_z {\bf v} \big) = -\frac{\partial}{\partial \phi} P -\rho \frac{\partial}{\partial \phi} U,
\ee
where $j_z = \rho r \sin\theta v_\phi$ is the $z$-component of the
angular momentum density. We integrate equation \ref{am1} over the
perturbed volume $\tilde{V}$ corresponding to the spherical shell 
initially at radius $r$ in the unperturbed star. With the Reynold's transport theorem this yields
\be
\label{am3}
\frac{d}{dt} \int d \tilde{V} j_z = - \int d \tilde{V} \frac{\partial}{\partial \phi} P - \int d \tilde{V} \rho \frac{\partial}{\partial \phi} U .
\ee
Note that the perturbed volume $d\tilde{V}$ can be expressed as
the unperturbed volume $dV$ plus a surface perturbation, i.e., 
$d\tilde{V} = dV + r^2 \xi_r dA$, where $dA$ is the solid angle element.
Then equation \ref{am3} becomes, to second order in perturbation amplitude,
\be
\label{am4}
\frac{d}{dt} J_z = - r^2 \int dA \xi_r \frac{\partial}{\partial \phi} \big( \delta P + \rho U) - \int dV \delta \rho \frac{\partial}{\partial \phi} U,
\ee
where we have ignored first order terms
that time average to zero. The left side of equation \ref{am4} is the
time derivative of the total angular momentum $J_z = \int d \tilde{V} j_z$ , 
while the first term on the right hand side is the angular
momentum flux term given in equation \ref{angmom} and the last term is
the torque from the tidal potential. The time derivative of the
angular momentum of any fluid element is zero if there is no wave
dissipation within the volume (Goldreich \& Nicholson 1989a) as shown
explicitly in Section \ref{dep}. Since the wave dissipation occurs
outside of our integration region, the terms on the right hand side of
equation \ref{am4} exactly cancel each other out. The torque is identical
in the rotating and non-rotating frames, and so the flux term in
equation \ref{am4} can be applied in the rotating frame (so long as
the change in wave frequency is taken into account). Equation
\ref{am4} can also be derived in the rotating frame, although the
derivation is more complex because torques from the Coriolis force
must be included.

It is also instructive to derive the energy flux through a comoving
fluid surface. In the rotating frame, we have
\be
\label{e1}
\frac{d}{dt} \int d \tilde{V} \varepsilon = - \int d \tilde{V} \rho \bnab \Phi \cdot {\bf v} - \int P {\bf v} \cdot d \tilde{{\bf S}},
\ee
where $\varepsilon$ is the energy density, $\Phi$ is the total
gravitational potential, and $\tilde{{\bf S}}$ is the perturbed
surface element. To second order,
\be
\label{e2}
\frac{d}{dt} \int d \tilde{V} \varepsilon = - \int d V g \ \delta \rho \ \delta v_r  - \int d V \rho \bnab U \cdot \delta {\bf v} - r^2 \int dA \ \delta P \ \delta v_r.
\ee
The term proportional to $g$ must vanish because the surface integral
$\int dA \ \delta \rho \ \delta v_r$ must vanish in order to
conserve mass.\footnote{This term must vanish to conserve mass,
  although, as pointed out in the text, it does not vanish until one
  includes the effect of a current that develops to counteract the
  Stokes drift current. Here we assume that such a current is indeed
  present, conserving mass and canceling the gravitational potential
  energy flux term of equation \ref{e2}.} We transform the second term
by noting that
\begin{align}
\label{urho}
\int d V \rho \bnab U \cdot \delta {\bf v} &= \int d V \bnab \cdot \big(\rho U \delta {\bf v} \big) - \int d V U \bnab \cdot \big(\rho \delta {\bf v} \big) \nonumber \\ 
 &= r^2 \int dA \rho U \delta v_r + \int d V U \frac{\partial}{\partial t} \delta \rho,
\end{align}
where the terms in the second line arise from applying Gauss's theorem and
the continuity equation, respectively. We then have
\be
\label{e3}
\frac{d}{dt} E =  -r^2 \int dA \big(\delta P + \rho U \big) \delta v_r -\int d V U \frac{\partial}{\partial t} \delta \rho.
\ee
The first term on the right hand side is the energy flux out the
surface, while the second is the work done by the tidal
potential. Recalling that $\delta v_r = - i \omega \xi_r$, it is
evident that the energy flux and angular momentum flux are related by
$\dot{J}_z = m \dot{E}/\omega$.

\subsection{Comparison to Previous Results}

We comment on the expressions for angular momentum fluxe used in
previous works such as Pantillon et al. (2007) and Mathis et al. (2009). 
These works quote the $z$-component of the angular momentum
flux as
\begin{align}
\label{mathis}
\dot{J}_z &= \int dA \rho r^3 \Big[ \sin \theta \delta v_r \delta v_\phi + 2 \sin \theta \cos\theta \Omega_s \xi_r \delta v_\theta \Big] \nonumber \\
 &= \frac{2 m'}{\omega} \dot{E}_k,
\end{align}
where $m'$ is a numerically calculated effective azimuthal wavenumber
that depends on the stellar spin frequency, and $\dot{E}_k$ is the
kinetic energy flux carried by the wave.

Equation \ref{mathis} can be somewhat reconciled with equations \ref{angmom} and \ref{enang}. We note that equation \ref{xip1} can be written as
\be
\label{phiacc}
\rho r \big(-i \omega \sin \theta \delta v_\phi + 2 \sin \theta \cos \theta \Omega_s \delta v_\theta \big) = - \frac{\partial}{\partial \phi} \Big( \delta P + \rho U \Big) - 2 \rho r \sin^2 \theta \Omega_s \delta v_r.
\ee
Multiplying by $\xi_r$ yields
\be
\label{phiacc2}
\rho r \big( \sin \theta \delta v_r \delta v_\phi + 2 \sin \theta \cos \theta \Omega_s \xi_r \delta v_\theta \big) = - \xi_r \frac{\partial}{\partial \phi} \Big( \delta P + \rho U \Big) - 2 \rho r \sin^2 \theta \Omega_s \xi_r \delta v_r.
\ee
The last term on the right hand side vanishes upon addition of its complex conjugate. Then we have
\begin{align}
\label{ammath}
\dot{J}_z &= - r^2 \int dA \xi_r \frac{\partial}{\partial \phi} \big( \delta P + \rho U) \nonumber \\
		&= \frac{m}{\omega} \dot{E},
\end{align}
in accordance with equations \ref{angmom} and \ref{enang}. Thus, it is
not necessary to numerically calculate the angular integral $m'$, and
the energy and angular momentum fluxes are always related by a factor
of $m$ as long as the unperturbed stellar structure is
axisymmetric. The factor of 2 difference between the energy flux in
equation \ref{ammath} and \ref{mathis} arises because equation
\ref{mathis} refers only to the kinetic energy flux of the wave,
$\dot{E}_k$, which is related to the total energy (kinetic and
potential) flux by $\dot{E}=2 \dot{E}_k$.

The conclusions above are independent of the traditional
approximation, although one may reach the same conclusion by
manipulating equations \ref{xitang} and \ref{xipang}. From equation
\ref{mathis}, using the traditional approximation, we have
\begin{align}
\label{amhough}
\dot{J}_z &= \int dA \rho r^3 \Big[ \sin \theta \delta v_r \delta v_\phi + 2 \sin \theta \cos\theta \Omega_s \xi_r \delta v_\theta \Big] \nonumber \\
	&= 2 \omega^2 \int dA \rho r^3 \Big[ \sin \theta {\rm Re}\Big( \xi_r^* \xi_\phi \Big)+ q \sin \theta \cos \theta {\rm Re}\Big( i \xi_r^* \xi_\theta \Big) \Big] \nonumber \\
	&= 2 \omega^2 \int dA \rho r^3 {\rm Re}\bigg[ \xi_r^* \bigg( \frac{\sin \theta}{1 - q^2 \cos^2 \theta} \Big[ \frac{i m }{\sin \theta} - i q \cos \theta \frac{\partial}{\partial \theta} \Big] \xi_\perp + \frac{i q \sin \theta \cos \theta}{1 - q^2 \cos^2 \theta} \Big[ \frac{\partial}{\partial \theta} - m q  \frac{\cos \theta}{\sin \theta} \Big] \xi_\perp \bigg) \bigg] \nonumber \\
	&= 2 \omega^2 \int dA \rho r^3 {\rm Re}\bigg[ \frac{1}{1 - q^2 \cos^2 \theta} \xi_r^* \Big( i m \big[ 1 - q^2 \cos^2 \theta \big] \Big) \xi_\perp \bigg]\nonumber \\
	 &= 2 m \omega^2 \rho r^3 {\rm Re} \bigg[ i \xi_r^* \xi_\perp \bigg].
\end{align}
The second line follows from $2 \Omega_s = \omega q$ and using $\delta {\bf v} = - i \omega \bxi$, the third from equations \ref{xitang} and \ref{xipang}, the fourth from combination of terms, and the fifth from angular integration over a sphere. Thus the traditional approximation obeys the angular momentum flux and energy flux relation of equation \ref{enang}.

Some previous authors (e.g., Lee \& Saio 1993) have obtained an
additional angular momentum flux term,
\be
\label{false}
\dot{J}_{z,\rho} = \Omega_s r^4 \int dA \sin^2 \theta \ \delta \rho \ \delta v_r.
\ee
This term does not actually contribute to angular momentum transfer via waves. It vanishes upon subtraction of terms due to Stokes drift velocities, as described in Section \ref{stokesdrift}. Regardless, in WD interiors the angular momentum flux from equation \ref{false} is small compared to that of equation \ref{angmom} because $|\delta \rho /\rho| \ll |\xi_\perp /r|$ for gravity waves in the WKB limit.

\subsection{Deposition of Angular Momentum}
\label{dep}

In this subsection we show explicitly how the angular momentum of a
shell of thickness $dr$ evolves with time. In particular, we show that
in the absence of local wave dissipation, there is no net torque on
the shell.

The amount of angular momentum deposited in a shell of thickness $dr$
due to the change in angular momentum flux across it is
\begin{align}
\label{dangmom}
\frac{\partial}{\partial t} 4 \pi r^2 j_z(r) dr &= dr \frac{\partial}{\partial r} \dot{J}_z(r) \nonumber \\
&=  2 dr \frac{\partial}{\partial r} \int dA {\rm Re} \bigg[ r^2 \xi_r^*(r,\theta,\phi) \frac{\partial}{\partial \phi} \bigg(\delta P(r,\theta,\phi) + \rho U(r,\theta,\phi) \bigg) \bigg] \nonumber \\ 
&=  2m dr \int dA {\rm Re} \bigg[ i  \bigg(\delta P(r,\theta,\phi) + \rho U(r,\theta,\phi) \bigg) \frac{\partial}{\partial r} \bigg( r^2 \xi_r^*(r,\theta,\phi) \bigg) \bigg] \nonumber \\ 
&+  2m dr \int dA {\rm Re} \bigg[ i \bigg( r^2 \xi_r^*(r,\theta,\phi) \bigg) \frac{\partial}{\partial r} \bigg(\delta P(r,\theta,\phi) + \rho U(r,\theta,\phi) \bigg) \bigg],
\end{align}
with the second line following by inserting equation \ref{angmom}. To
evaluate the radial derivatives, we use equations \ref{xir1} and
\ref{cont}, along with the adiabatic relation of equation \ref{drho}
to find
\begin{align}
\label{dangmom2}
\frac{\partial}{\partial t} 4 \pi r^2 j_z(r) dr &= 2 m dr \int dA {\rm Re} \bigg[ - i r^2 U \delta \rho^* - i r^2 \big( \delta P + \rho U \big) \bnab_\perp \cdot \bxi_\perp^* + 2 \Omega_s \omega \rho r^2 \sin \theta \xi_r^* \xi_\phi \bigg] \nonumber \\ 
&= 2 m dr \int dA {\rm Re} \bigg[ - i r^2 U \delta \rho^* - \bnab_\perp \cdot \Big[ i r^2 \big( \delta P + \rho U \big) \bxi_\perp^* \Big] + \bxi_\perp^* \cdot \bnab_\perp \Big[ i r^2 \big( \delta P + \rho U \big) \Big] + 2 \Omega_s \omega \rho r^2 \sin \theta \xi_r^* \xi_\phi \bigg] \nonumber \\
&= 2 m dr \int dA {\rm Re} \bigg[ - i r^2 U \delta \rho^*  - 2 \Omega_s \omega \rho r^2 \cos \theta \big( \xi_\theta \xi_\phi^* - \xi_\theta \xi_\phi^* \big) - 2 \Omega_s \omega \rho r^2 \cos \theta \big(\xi_r \xi_\phi^* - \xi_r^* \xi_\phi \big) \bigg] \nonumber \\
&= 2 m dr \int dA {\rm Re} \bigg[ - i r^2 U \delta \rho^* \bigg] .
\end{align}
The second line arises from rearranging the horizontal divergence
term. The surface integral of a horizontal divergence vanishes on a
closed surface, eliminating the second term on the second line, and
inserting equations \ref{xit1} and \ref{xip1} we obtain the third
line. The terms in parentheses are purely imaginary, leaving us with
the simple result of the fourth line.

The fourth line of equation \ref{dangmom2} is equal to the negative of
the torque exerted on the shell by the tidal potential. Therefore, in
the absence of local non-adiabatic effects, the angular momentum of
any region of the star remains unchanged by tidal torques and wave
propagation, as proven more generally in Golreich \& Nicholson
1989a. Basically, in the absence of dissipation, any angular momentum
deposited in a region via tidal torques is transported away by the
waves.

This fact can help us understand why we need only calculate the
dynamical tide contributions to equation \ref{angmom}. Although
the equilibrium tide-dynamical tide cross terms contribute to the angular
momentum flux, they are exactly canceled by torques from the tidal
potential. The terms due purely to dynamical tide 
cannot be ignored because, away from the regions of excitation or damping, their
angular momentum luminosity remains constant (and therefore do not
contribute to equation \ref{dangmom2}). This assertion is easily
verified in the WKB limit (see Paper II). To calculate the total
angular momentum luminosity flowing into the outer regions of the
star, one could either evaluate the surface integral of equation
\ref{angmom}, or the volume integral of equation \ref{am4}. However,
since the integrand of the last term equation \ref{am4} is highly
oscillatory with radius, a volume integral is highly susceptible to
numerical errors, and we find it much easier to use the last line of
equation \ref{angmom}.

\section{Inner Boundary Condition}
\label{innerbc}

The traditional approximation encounters problems at the center of the
star where $r \rightarrow 0$ because both postulates of the
approximation are broken (near the center of the star $N \rightarrow
0$ and $\xi_r \approx \xi_\theta$). As we will see below, the
manifestation of these problems is the divergence of the dynamical
tide solution at the center of the star, which is obviously
unphysical.

The oscillation equations \ref{dp} and \ref{dxir} take the form at the
center of the star (where $r \rightarrow 0$, $\rho \rightarrow
\rho_c$, $g \rightarrow 4 \pi G \rho_c r/3$, and $\partial g /\partial
r \rightarrow 4 \pi G \rho_c /3$, with $\rho_c$ the central density):
\begin{align}
& \frac{\partial}{\partial r} \delta P^{\rm dyn} \simeq \rho \omega^2  \xi_r^{\rm dyn} + h_{klm} \rho \omega^2 \xi_r^{\rm eq} + h_{klm} m q  \rho \omega^2 \xi_\perp^{\rm eq},\label{dpa}\\
&\frac{\partial }{\partial r} \xi_r^{\rm dyn} \simeq  - \frac{2}{r} \xi_r^{\rm dyn} + \frac{\lambda_k}{\rho r^2 \omega^2} \delta P^{\rm dyn} - \frac{h_{klm}l(l+1)}{r} \xi_\perp^{\rm eq}  - \frac{g_{klm}}{r} \xi_r^{\rm eq}.
\label{dxira}
\end{align}
Imposing a Frobenius expansion such that $\xi_r^{\rm dyn} \propto r^\alpha$, $\delta P^{\rm dyn} \propto r^{\alpha+1}$, and defining $U_c = - \frac{3 M' W_{22}}{4 \pi \rho_c a^{l+1}}$, we have
\be
\label{dpa1}
\frac{\alpha +1}{\rho r \omega^2} \delta P^{\rm dyn} \simeq \xi_r^{\rm dyn} +  h_{klm} U_c r^{l-1} + \frac{h_{klm} m q}{l} U_c r^{l-1},
\ee
\be
\label{dxira1}
\frac{\alpha + 2}{r} \xi_r^{\rm dyn} \simeq  \frac{\lambda_k}{\rho r^2 \omega^2} \delta P^{\rm dyn} - (l+1) h_{klm} U_c r^{l-2} - g_{klm} U_c r^{l-2}.
\ee
These equations have both particular (forced) solutions and homogeneous (free) solutions. The forced solutions have $\xi_r^{\rm dyn} \propto r^{l-1}$ and thus vanish near the origin. The free solution satisfies 
\be
\label{dpa2}
\frac{\alpha +1}{\rho r \omega^2} \delta P^{\rm dyn} \simeq \xi_r^{\rm dyn},
\ee
\be
\label{dxira2}
\frac{\alpha + 2}{r} \xi_r^{\rm dyn} \simeq  \frac{\lambda_k}{\rho r^2 \omega^2} \delta P^{\rm dyn},
\ee
which requires 
\be
\label{alpha}
\alpha^2 + 3 \alpha + (2 - \lambda_k) = 0.
\ee
The solutions of this quadratic equation are 
\be
\label{alpha2}
\alpha = -\frac{3}{2} \pm \frac{1}{2} \sqrt{9 + 4(\lambda_k -2) }.
\ee
If $\lambda_k > 2$, there exists a regular solution with $\alpha > 0$,
allowing one to set $\xi_r^{\rm dyn} = \delta P^{\rm dyn} = 0$ at
$r=0$. However, if $\lambda_k < 2$ there is no solution for which
$\alpha > 0$, and thus there exists no regular solution at $r=0$. Any
solution of equations \ref{dp} and \ref{dxir} for $\lambda_k < 2$
therefore diverges near the origin.

As stated above, this unphysical divergence can be attributed to the
breakdown of the traditional approximation near the origin. This issue
is especially worrisome for the $k=-2$, $m=-2$ r-waves for which
$\lambda_k <2$ for all values of $q$. In practice, when computing a
solution with $\lambda_k < 2$, we choose the positive solution of
equation \ref{alpha2} and use for our inner boundary condition the
relation
\be
\label{inbc}
\rho r_{\rm in} \omega^2 \xi_r^{\rm dyn} = (\alpha + 1) \delta P^{\rm dyn},
\ee
computed at some very small vale of $r_{\rm in}$. We find that the
choice of $r_{\rm in}$ does not affect the solution near the outer
boundary, nor does it affect computed values of
$F(\omega,q)$. Although this method cannot be used to characterize the
dynamical tide solution near the origin, we conclude that it yields
accurate solutions away from the origin where the traditional
approximation is valid.

\def\apj{{Astrophys. J.}}
\def\apjs{{Astrophys. J. Supp.}}
\def\mnras{{Mon. Not. R. Astr. Soc.}}
\def\prl{{Phys. Rev. Lett.}}
\def\prd{{Phys. Rev. D}}
\def\apjl{{Astrophys. J. Let.}}
\def\pasp{{Publ. Astr. Soc. Pacific}}
\def\aapr{{Astr. Astr. Rev.}}

%%%%%%%%%%%%%%%%%%%%%%


\begin{thebibliography}{99}

\bibitem[]{}
Alexander, M.E. 1973, Astrophys. Space Sci., 23, 459

\bibitem[]{}
Andrews, D., McIntyre, M., 1978, JFM, 89, 609

\bibitem[]{}
Andrews, D., McIntyre, M., 1978, JFM, 89, 647

\bibitem[]{}
Bildsten, L., Ushomirsky, G., Cutler, C., 1996, ApJ, 460, 827

\bibitem[]{}
Bloom, J., et al., 2012, ApJL, 744, 17

\bibitem[]{}
Brown, W., Kilic, M., Hermes, J.J., Allende Prieto, C., Kenyon, S.J., Winget, D.E. 2011, ApJ, 737, 23

\bibitem[]{}
Brown, W., Kilic, M., Allende Prieto, C., Kenyon, S., 2012, ApJ, 744, 142

\bibitem[]{}
Burkart, J., Quataert, E., Arras, P., Weinberg, N., 2013, MNRAS, 433, 332

\bibitem[]{}
Chapman, S., Lindzen, R.S., 1970, Atmospheric Tides, Dordrecht: Reidel

\bibitem[]{}
Clayton, G., 2012, JAVSO, 40, 539

\bibitem[]{}
Dall'Osso, S., Rossi, E.M., 2013,  arXiv:1308.1664

\bibitem[]{}
Dan, M., Rosswog, S., Guillochon, J., Ramirez-Ruiz, E. 2012, MNRAS, 422, 2417

\bibitem[]{}
Dan, M., Rosswog, S., Brueggen, M., Podsiadlowski, P., 2013, arXiv:1308.1667

\bibitem[]{}
Dintrans, B., Rieutord, M., Valdetarro, L., 1999, JFM, 398, 271

\bibitem[]{}
Dintrans, B., Rieutord, M., 2000, A\&A, 354, 86

\bibitem[]{}
Di Stefano, R. 2010, ApJ, 719, 474

\bibitem[]{}
Fuller, J., Lai, D. 2011, MNRAS, 412, 1331 (Paper I)

\bibitem[]{}
Fuller, J., Lai, D., 2012, MNRAS, 421, 426 (Paper II)

\bibitem[]{}
Fuller, J., Lai, D., 2012, ApJL, 756, 17 (Paper III)

\bibitem[]{}
Fuller, J., Lai, D., 2013, MNRAS, 430, 274 (Paper IV)

\bibitem[]{}
Gilfanov, M., Bogdan, A., 2010, Nature, 463, 924

\bibitem[]{}
Goldreich, P., Nicholson, P., 1989, ApJ, 342, 1075

\bibitem[]{}
Goldreich, P., Nicholson, P., 1989, ApJ, 342, 1079

\bibitem[]{}
Golreich, P., \& Soter, S. 1966, Icarus, 5, 375

\bibitem[]{}
Gonzalez Hernandez, J., et al. 2012, Nature, 489, 533

\bibitem[]{}
Han, Z., Webbink, R., 1999, A\&A, 349, L17

\bibitem[]{}
Han, Z., Podsiadlowski, Ph., Maxted, P., Marsh, R., Ivanova, N., 2002, MNRAS, 336, 449

\bibitem[]{}
Heber, U., 2009, Ann. Rev. Astron. \& Astroph., 47, 211

\bibitem[]{}
Hut, P. 1981, A\&A, 99, 126

\bibitem[]{}
Iben, I., Tutukov, A. 1984, ApJS, 54, 335

\bibitem[]{}
Iben, I., Tutukov, A., Yungelson, L., ApJ, 1996, 456, 750

\bibitem[]{}
Iben, I., Tutukov, A., Fedorova, A. 1998, ApJ, 503, 344

\bibitem[]{}
Jeffery, C. Karakas, A., Saio, H., 2011, MNRAS, 414, 3599

\bibitem[]{}
Kilic, M. Brown, W., Allende Prieto, C., Kenyon, S., Heinke, C., Agüeros, M., Kleinman, S., 2012, ApJ, 751, 141

\bibitem[]{}
Kilic, M., et al., 2013, arXiv:1310.6359

\bibitem[]{}
Kulkarni, S.R., van Kerkwijk, M.H., 2010, ApJ, 719, 1123

\bibitem[]{}
Kumar, P., Talon, S., Zahn, J.P., 1999, ApJ, 520, 859

\bibitem[]{}
Lee, U., Saio, H., 1993, MNRAS, 261, 415

\bibitem[]{}
Lee, U., Saio, H., 1997, ApJ, 491, 839

\bibitem[]{}
Li, W., et al., 2011, Nature, 480, 348

\bibitem[]{}
Loren-Aguilar, P., Isern, J., Garcia-Berro, E. 2009, AA, 500, 1193

\bibitem[]{}
Maoz, D., Sharon, K., Gal-Yam, A. 2010, ApJ, 722, 1979

\bibitem[]{}
Marsh, T., Nelemans, G., Steeghs, D., 2004, MNRAS, 350, 113

\bibitem[]{}
Marsh, T., 2011, CQGra, 28, 094019

\bibitem[]{}
Mathis, S., Talon, S., Pantillon, F.P., Zahn J.P., 2008, SoPh, 251, 101

\bibitem[]{}
Mathis, S., 2009, A\&A, 506, 811

\bibitem[]{}
Mullally, F., Badenes, C., Thompson, S.E., Lupton, R. 2009, ApJ, 707, L51

\bibitem[]{}
Nelemans, G., Steeghs, D., Groot, P., 2001, MNRAS, 326, 621

\bibitem[]{}
Nelemans, G., Portegies Zwart, S.F., Verbunt, F., Yungelson, L.R., 2001, A\&A, 368, 939

\bibitem[]{}
Nomoto, K., Kondo, Y., 1991, ApJL, 367, 19

\bibitem[]{}
Pantillon, F.P., Talon, S., Charbonnel, C., 2007, A\&A, 474, 155

\bibitem[]{}
Piro, T. 2011, ApJ, 740, L53

\bibitem[]{}
Raskin, C., Scannapieco, E., Fryer, C., Rockefeller, G., Timmes, F. 2012, ApJ, 746, 62

\bibitem[]{}
Saoi, H., Nomoto, K., 1985, A\&A, 150, 21

\bibitem[]{}
Saio, H., Jeffery, C., 2000, MNRAS, 313, 671

\bibitem[]{}
Saio,, H., Nomoto, K., 2004, ApJ, 615, 444

\bibitem[]{}
Schaefer, B., Pagnotta, A., 2012, Nature, 481, 164

\bibitem[]{}
Segretain, L., Chabrier, G., Mochkovitch, R. 1997, ApJ, 481, 355

\bibitem[]{}
Steinfadt, J., Kaplan, D.L., Shporer, A., Bildsten, L., Howell, S.B. 2010, ApJ, 716, L146

\bibitem[]{}
Townsend, R.H.D., 2003, MNRAS, 340, 1020

\bibitem[]{}
Tutukov, A., Yungelson, L., 1996, MNRAS, 280, 1035

\bibitem[]{}
Valsecchi, F., Farr, W., Willems, B., Kalogera, V., 2012, arXiv:1210.5023

\bibitem[]{}
Van Kerkwijk, M.H., Chang, P., Justham, S. 2010, ApJL, 722, 157

\bibitem[]{}
Warner, B., 1995, Ap\&SS, 225, 249

\bibitem[]{}
Webbink, R.F. 1984, ApJ, 277, 355

\bibitem[]{}
Willems, B., Deloye, C.J., Kalogera, V. 2010, ApJ, 713, 239

\bibitem[]{}
Yoon, S.-C., Podsiadlowski, Ph., Rosswog, S., 2007, MNRAS, 380, 933

\end{thebibliography}
\end{document}